\documentclass[preprint,12pt]{elsarticle}

\usepackage[english]{babel}
\usepackage{amssymb}
\usepackage{pifont}
\usepackage{natbib}
\usepackage[margin=17mm]{geometry}
\usepackage{fleqn}
\usepackage{graphicx}
\usepackage{caption}
\usepackage{subcaption}
\usepackage{hyperref}
\usepackage{color}
\addto\captionsenglish{}

\usepackage[cmex10]{amsmath}
\usepackage{amssymb}
\usepackage{booktabs}

\graphicspath{ {./Figures/} }

\newcommand{\beq}{\begin{equation}}
\newcommand{\eeq}{\end{equation}}
\newcommand{\bea}{\begin{eqnarray}}
\newcommand{\eea}{\end{eqnarray}}
\newcommand\sect[1]{Section~\ref{sec:#1}}
\newcommand\app[1]{\ref{app:#1}}
\newcommand\fig[1]{Figure~\ref{fig:#1}}
\newcommand\tab[1]{Table \ref{tab:#1}}
\newcommand\eqn[1]{Eq.~(\ref{eq:#1})}

\newcommand{\red}[1]{\textcolor{black}{#1}}
\newcommand{\blue}[1]{\textcolor{black}{#1}}

\journal{Mechanical Systems and Signal Processing}

\begin{document}

\begin{frontmatter}

\title{Dictionary learning approach to monitoring of wind turbine drivetrain bearings}

\author[srt,utc]{S.~Martin-del-Campo}
\ead{sergio.martindelcampo@ltu.se}
\author[srt]{F.~Sandin}
\ead{fredrik.sandin@ltu.se}
\author[tvm,utc]{D.~Str\"{o}mbergsson}
\ead{daniel.strombergsson@ltu.se}

\address[srt]{EISLAB, Lule{\aa} University of Technology (LTU)}
\address[tvm]{Div. of Machine Elements, Lule{\aa} University of Technology (LTU)}
\address[utc]{SKF-LTU University Technology Center}

\begin{abstract}
Condition monitoring is central to the efficient operation of wind farms due to the challenging operating conditions, rapid technology development and \blue{large} number of aging wind turbines.
In particular, predictive maintenance planning requires \blue{the} early detection of faults with few false positives.
\blue{Achieving this type of detection} is a challenging problem due to the complex and weak signatures of some faults, \blue{particularly the faults that occur} in some of the drivetrain bearings.
Here, we investigate recently proposed condition monitoring methods based on unsupervised dictionary learning using vibration data recorded
over 46 months under typical industrial operations\blue{. Thus, we contribute} novel test results and \blue{real--world data that are} made publicly available.
\blue{The results} of former studies addressing condition--monitoring tasks using dictionary learning indicate that unsupervised feature learning is useful for diagnosis and anomaly detection purposes.
However, these studies are based on small sets of labeled data from test rigs operating under controlled conditions that focus on classification tasks, which are useful for quantitative method comparisons but gives little \blue{insight into} how useful these approaches are in practice.
In this study, dictionaries are learned from gearbox vibrations in six different turbines, 
and the dictionaries are subsequently propagated over a few years of monitoring data when faults are known to occur.
We perform the experiment using two different sparse coding algorithms to investigate if the algorithm selected affects the features of abnormal conditions.
We calculate the dictionary distance between the initial and propagated dictionaries and find \blue{the} time periods of abnormal dictionary adaptation starting six months before a drivetrain bearing replacement and one year before the resulting gearbox replacement.
\blue{In addition, we} investigate the distance between dictionaries learned from geographically \blue{close} turbines of the same type \blue{under} healthy conditions.
\blue{We} find that the features learned are similar and that a dictionary learned from one turbine can be useful for monitoring \blue{a} similar turbine.
\end{abstract}


\end{frontmatter}

\section{Introduction}

Wind power is a renewable energy source that \blue{is growing} rapidly and provides more than 11\% of the electrical power in the European Union \cite{GWEC2017}.
Wind power is harvested by wind farms, which typically include many similar wind turbines.
Wind turbines are based on relatively new technology \blue{that} has been scaled up from \blue{approximately} 2MW to 10MW per turbine in one decade.
The rapid development in combination with the challenging operating conditions of wind turbines over the typical 20-year service life implies that condition monitoring and predictive maintenance are central issues.
When maintenance is needed, the cost of crane mobilization and energy production losses are high, and there are challenges \blue{acquiring} spare parts in this rapidly expanding industry.
The gearbox is a major component of a wind turbine, and the rolling element bearings that support the rotating components in the drivetrain are essential for reliable operation.
Monitoring these bearings is an important and challenging issue given the predominance of bearing faults in wind turbines \cite{Crowther2011} and the complex and weak signatures of some faults~\cite{Liu2017}.
\blue{Because nearby} turbines face similar environmental and operational conditions, methods can be adapted and validated with data from multiple machines.

Condition-based maintenance requires continuous monitoring of the machine to detect incipient failures so that the maintenance actions can be scheduled \blue{efficiently}~\cite{Hossain2018}.
This procedure involves three stages: data acquisition, feature extraction and diagnostics identification.
The principle behind identification is that a ``significant change [of a feature] is indicative of a developing failure''~\cite{GarciaMarquez2012}.
Feature selection and extraction is a key problem that typically determines the performance of decision support functions and thereby the efficiency of the condition monitoring system.
A feature is an individual measurable signal property or pattern that is characteristic of some particular type of source.
The condition monitoring of wind turbines \blue{typically} uses methods based on feature extraction with signals originating from vibration sensors mounted on the drivetrain.
Hossain~\cite{Hossain2018} describes common faults in wind turbines and the typical features used in their identification.

The features used in the diagnostics of a wind turbine can be classified in three categories: time domain, frequency domain and joint time-frequency domain.
Time--domain features include traditional statistical parameters \blue{such as the} root mean square \blue{(RMS)}, crest factor and kurtosis.
\blue{The trends of these} parameters are used as indicators of the deterioration of the machine \cite{Lacey2008}.
Frequency--domain features are typically derived by \blue{the} conversion of the time-domain signal to the frequency domain using the fast Fourier transform (FFT).
Kinematic \blue{data such as} bearing defect frequencies are used to extract information in selected frequency bands \cite{Jardine2006}, \blue{and thereby form} a smaller set of features that can be trended and monitored.
Analysis methods based on joint time-frequency domain features are also used, but \blue{these} are more recent developments compared to the \blue{time--domain} and frequency--domain methods.
The wavelet transform is one example, which is useful for the analysis of nonstationary signals \cite{GarciaMarquez2012}.
However, \blue{to date,} such methods have not been widely used in industry because the analysis is more complex and requires trained experts \blue{to interpret} the results~\cite{Tse2011}.
Further information about these analysis methods can be found in, \blue{for example,} \cite{Dias2016,Rai2016}.

Features are typically manually selected by experts, which implies that the features are selected without explicit knowledge \blue{of} the state of each \blue{individual} machine.
Furthermore, the dependence on experts is a bottleneck that limits the scalability of condition monitoring systems.
\blue{In general, the detection}, prediction and diagnosis of faults in a rolling element bearing are challenging tasks due to the high number of variables \blue{that affect} the operation.
Thus, a machine learning approach can be useful in the development of more automated diagnosis and prognosis systems.
Supervised machine learning is one approach\blue{, but it requires} labeled data for training, which \blue{are} difficult and expensive to generate \cite{Randall2011}.
An alternate approach is unsupervised learning methods, which, for example, can be used for feature learning and anomaly detection purposes.

Here, we investigate an online feature learning approach based on dictionary learning \blue{that enables the} optimization of the monitored feature set \blue{for} each individual machine.
In particular, we use dictionary learning to study signals recorded from vibration sensors installed on gearboxes in 2.5~MW turbines at a wind farm in northern Sweden.
The learned features define a set of overcomplete and shift-invariant waveforms \blue{that} are used to determine a sparse approximation of the corresponding vibration signal.
We are interested in measures that can be used to track \blue{the} changes of such waveforms over time, \blue{as they} can be useful for the detection of abnormal changes.

Dictionary learning~\cite{Mallat2008} and convolutional sparse coding~\cite{Papyan2018} has attracted broad interest.
Variations of the dictionary learning method have successfully been used in tasks such as signal compression, detection, separation and denoising \citep{elad2010sparse,Bruckstein2009,Starck2010}.
The methods developed here are based on the work by Smith and Lewicki \cite{Smith2005,Smith2006}, which is inspired by \blue{the} earlier work of Olshausen and Field \cite{Olshausen96,Olshausen97} in the area of sparse visual coding.
The methodology includes a sparse regularization mechanism that \red{reduces} the influence of noise and some of the redundancy \red{that is} typically present in raw sensor signals.
\blue{Here, the} hypothesis is that the same general approach can be used to characterize and analyze the signals generated by a rotating machine~\cite{MartinDelCampo2015}.

Liu et al.~\cite{Liu2011} were the first to apply dictionary learning to a dataset with bearing vibration signals.
\blue{These authors} trained dictionaries of waveforms of fixed length for different bearing conditions.
The learned dictionaries were subsequently merged and used to classify the type of fault \blue{with} a linear classifier.
Furthermore, Martin-del-Campo et al.~\cite{MartinDelCampo2013} showed that it is possible to distinguish different operational conditions through \blue{the} learning of shift-invariant waveforms where the lengths of the waveforms are also optimized.
Chen et al.~\cite{Chen2014b} use a dictionary learning approach to detect a fault in a gearbox by \blue{the} identification of impulse-like components in a vibration signal.
Tang et al.~\cite{Tang2014} \blue{used} shift-invariant sparse coding to generate a set of latent components that act as fault filters in a bearing or a gearbox.
\blue{Moreover,} studies by Ahmed et al.~\cite{Ahmed2018} and He et al.~\cite{He2018} \blue{proposed} classification strategies that use the learned sparse representations on stacked autoencoders and large memory storage and retrieval neural networks, respectively. 
Further extensions of the work by Liu et al.~\cite{Liu2011} \blue{had} been developed by Wang et al.~\cite{Wang2015} and Zhou et al.~\cite{Zhou2016},
who used the same dictionary learning method with different classification strategies.

Studies of dictionary learning for fault detection with bearing signals are based on simulated data and/or data from controlled experiments, where the faults are artificially introduced.
Furthermore, most of these studies investigate \blue{either} how the learned atoms can be used \blue{to classify faults} or how to improve the accuracy of such a classifier.
Here, we extend the former studies with an investigation based on real--world vibration data collected from \blue{the} vibration sensors on gearboxes in multiple wind turbines over an extended period of nearly four years.
The data \blue{have been released publicly;} see \app{data} for further information. 
The output shaft bearing and, subsequently, the gearbox were replaced in one of the turbines considered in this study.
In addition to considering the fault detection problem, we investigate whether the dictionary of waveforms learned for one turbine is useful for the analysis of the corresponding signal in \blue{a} nearby turbine of the same type.
Furthermore, we study the possibility \blue{of using} dictionary propagation and dictionary--based indicators to identify bad actors in a population of wind turbines, in a similar way that trend analysis is currently used to monitor the conditions of turbines.
The experiments presented below also include a comparison of two different sparse coding algorithms.

The dictionary learning method and the proposed dictionary--based indicators are described in \sect{dictionary_learning_approach}.
The data used are described in \sect{datasource}, and the results are presented in \sect{results}, followed by a discussion of the results in \sect{discussion}.

\section{Dictionary learning method}
\label{sec:dictionary_learning_approach}

\subsection{Sparse signal model}

The signal $S(t)$ is modeled as a linear superposition of Gaussian noise and waveforms with compact support
\beq
    S(t) =  \sum_{i=1}^{N} a_{i} \phi_{m(i)}(t - \tau_{i}) + \epsilon(t).
    \label{eq:genmodel}
\eeq
The functions $\phi_m(t)$ are {\em atoms} that are learned from the signal, which we \blue{also} refer to as features.
\red{An {\em atom} is an elementary waveform that describes a recurring feature of the signal.}
A set of atoms $\phi_m(t)$ defines a dictionary $\Phi$ \blue{that consists} of \textit{M} atoms
\beq
\Phi = \left\{ \phi_1, \cdots, \phi_{M} \right\}.
\eeq
\red{The triple $m(i),\tau_{i},a_{i}$ defines one atom instance.
We refer to the ratio between the total number of atom instances, $N$, and the total number of signal (segment) samples as the sparsity level.
The temporal position and amplitude of the $i$-th instance of atom $\phi_m(t)$ are denoted by $\tau_{i}$ and $a_{i}$, respectively.
The term $\epsilon(t)$ represents the model residual, which includes Gaussian noise.}

The inverse problem defined by \eqn{genmodel} is solved with an iterative two-step optimization process for each consecutive signal segment:
\begin{enumerate}
\item[1.] {\it Sparse coding} -- \blue{While maintaining a fixed dictionary}, determine the parameters $m(i)$, $\tau_{i}$ and $a_{i}$ of the $N$ atom instances in \eqn{genmodel} using the Matching Pursuit~\cite{Mallat1993} or Orthogonal Matching Pursuit algorithms \cite{Pati1993}.
\item[2.] {\it Dictionary update} -- Given the set of atom instances and the residual $\epsilon(t)$, update the atoms in the dictionary $\Phi$ using a probabilistic gradient method (described below).
\end{enumerate}
Step 1 is a convolutional sparse coding~\cite{Szlam2010} process that is repeated until a stopping condition is reached.
Typically, the stopping condition is defined in terms of the total number of terms $N$ of the approximation, as in this case\blue{; alternatively, this condition} can be defined in terms of the approximation error or signal-to-residual ratio.
By defining the stopping condition in terms of the number of atom instances, the sparsity of the model in \eqn{genmodel} is directly related to the number of iterations of the optimization process\blue{; hence,} the computational requirements for online operation are well defined.
Step 2, the dictionary update, is performed iteratively after \blue{the} sparse coding of each signal segment.
The process continues until there are no more signal segments.
\blue{In this way, the} online learning of atoms by processing consecutive signal segments is possible \cite{MartinDelCampo2015}.

\subsection{Signal encoding algorithm}

The model described by \eqn{genmodel} describes a continuous signal $S(t)$ as a linear combination of atoms.
However, the problem \blue{of finding} the optimal linear combination of an overcomplete set of atoms is an intractable (NP hard) problem.
Therefore, several algorithms have been proposed to find approximate solutions to this problem.
One set of algorithms are greedy algorithms that rely on an iterative process to create the sparse representation, for example Matching Pursuit~\cite{Mallat1993}, Orthogonal Matching Pursuit~\cite{Mailhe2009} and Gradient Pursuit~\cite{Blumensath2008}.

Here, we use the Matching Pursuit (MP) algorithm and the Orthogonal Matching Pursuit (OMP) algorithm to obtain a sparse approximation of each signal segment.
Both algorithms are used to decompose the signal given a dictionary of atoms.
The algorithms operate on the residual of the signal, which initially is the signal segment to be decomposed: $R_{0}(t) = s_{k}(t)$.
The cross-correlation between the residual and all \blue{of} the elements of $\Phi$ is calculated in each iteration. 
An atom instance is defined by the atom with the maximum cross-correlation (inner product) across all possible timeshifts, and the amplitude $a_{i}$ is defined by
\beq
   a_{i} = \langle s_{k}(t) | \phi_{m(i)}(t - \tau_{i}) \rangle,
   \label{eq:MPdecomp}
\eeq
where the temporal position $\tau_{i}$ is
\beq
   \tau_{i} = \arg \max_{i}~\langle s_{k}(t) | \phi_{m(i)}(t - \tau_{i}) \rangle.
   \label{eq:MPindproj}
\eeq
This process is repeated by determining a new atom instance for each iteration until the stopping condition is fulfilled.
In each iteration, the atom instance $a_{i} \phi_{m(i)}(t - \tau_{i})$ is subtracted from the residual to form a new residual to be used in the next iteration.

The MP and OMP methods have different residual update rules.
In MP, the updated residual of the signal $R_{i}(t)$ after the \textit{i}-th iteration is given by
\beq
	R_{i}(t) = R_{i-1}(t) - a_{i} \phi_{m(i)}(t - \tau_{i}).
	\label{eq:MPResidualUpdate}
\eeq 
The OMP algorithm updates all coefficients $a_{i}$ with an orthogonal projection of the signal segment onto the set of all previously selected atoms.
Thus, the updated residual of the signal, $R_{i}(t)$, after the \textit{i}-th iteration is
\beq
	R_{i}(t) = R_{i-1}(t) - \Phi_{i} (\Phi_{i}^{T} \Phi_{i})^{-1} \Phi_{i}^{T} R_{i-1}(t),
	\label{eq:OMPResidualUpdate}
\eeq 
where $\Phi_{i}$ is the updated dictionary in the \textit{i}-th iteration.

The iterative process continues until the stop condition is reached.
\blue{This stop condition is defined based on} the number of terms $N$ relative to the number of samples of each signal segment $S(t)$. 
\blue{Moreover, the number} of terms also \blue{determines} the sparsity of the approximation.
For further details about the MP method used here, see~\cite{MartinDelCampo2013}; \blue{for} the OMP method, see \cite{Mailhe2009}.

\subsection{Learning of shift-invariant dictionary}

The next step in the iterative optimization process is to update the atoms in the dictionary $\Phi$ using the sparse approximation of the signal.
The goal is to optimize the set of atomic waveforms $\phi_m$ in the dictionary $\Phi$ to minimize the residual of the sparse approximation.
One solution to this problem can be obtained by rewriting \eqn{genmodel} in probabilistic form:
\bea
        p(s_{k} | \Phi)\, = \int p(s_{k} | a,\Phi) p(a) da\quad  \approx\,\, p(s_{k} | \hat{a},\Phi) p(\hat{a}), 
        \label{eq:probreprS}
\eea
where $\hat{a}$ is the maximum a posteriori (MAP) estimate of $a$ \cite{Smith2005},
\beq
        \hat{a} = \arg \max_{a} p(a | s_{k}, \Phi) = \arg \max_{a} p(s_{k} | a, \Phi )  p(a),
        \label{eq:inferS}
\eeq
that is generated by the matching pursuit or orthogonal matching pursuit algorithms \cite{Smith2006}.
Furthermore, we assume that the noise term $\epsilon$ in \eqn{genmodel} is Gaussian.
Therefore, the data likelihood $p(s_{k} | a,\Phi )$ is also Gaussian and takes the form
\beq
        p(s_{k} | a,\Phi )  \approx \exp\left(- \frac{\| s_{k}-a\Phi \|^2}{2\sigma_{\epsilon}^{2}}\right)                   ,
        \label{eq:likeli}
\eeq
where
\beq
        \| s_{k}-a\Phi \|^2 = \| s_{k} - \sum_{i=1}^{N} a_{i} \phi_{m(i)}(t - \tau_{i}) \|^2,
        \label{eq:likeliExp}
\eeq
and $\sigma_{\epsilon}^{2}$ is the variance of the noise.
Note that $s_{k}$, $a$ and $\Phi$ are matrices in these probabilistic expressions and that the dictionary $\Phi$ includes all \blue{of the} possible shifts of each atom $\phi_m$.

Under these assumptions, the atoms in the dictionary can be optimized by performing gradient ascent on the approximate log data probability,
thus, resulting in a gradient of \eqn{probreprS} \blue{of} the form
\beq
        \frac{\partial}{\partial \phi_m} \log (p(s_{k} | \Phi)) = \frac{1}{\sigma_{\epsilon}^{2}} \sum_i  a_{i} \left[ s_{k} - \hat{s_{k}} \right]_{\tau_{i}}.
        \label{eq:probformexp}
\eeq
The term $\left[ s_{k} - \hat{s_{k}} \right]_{\tau_{i}}$ represents the final residual of the sparse approximation (when the stop condition is met) that coincides with each atom at \blue{its respective temporal position} $\tau_{i}$ identified by MP or OMP.
\blue{Consequently}, the gradient of each atom in the dictionary is proportional to the sum of \blue{the} residuals within the support of each atom instance.

The use of the gradient for dictionary learning requires a step length parameter $\eta$ \blue{that} determines how much the atoms are updated.
The resulting update rule for an atom is
\beq
        \phi_m \rightarrow \phi_m
        + \frac{\eta}{\sigma_{\epsilon}^{2}} \sum_i  a_{i} \left[ s_{k} - \hat{s_{k}} \right]_{\tau_{i}}.    
        \label{eq:probreprSderSolvsimplr}
\eeq
\blue{Therefore,} the learning rate depends on how often atoms are selected during the sparse coding step,
which implies that the learning rate of atoms can be different and that some atoms may not learn at all
(see \cite{EMP2017} for an alternative dictionary learning method where this is not the case).
Furthermore, we zero-pad all atoms with ten elements at each tail
and allow an atom to grow in length if the RMS of the tail exceeds 0.1 of the atom RMS,
as described in \cite{Smith2006}.

\begin{figure}[!h]
\centering
\includegraphics[width=1.00\linewidth]{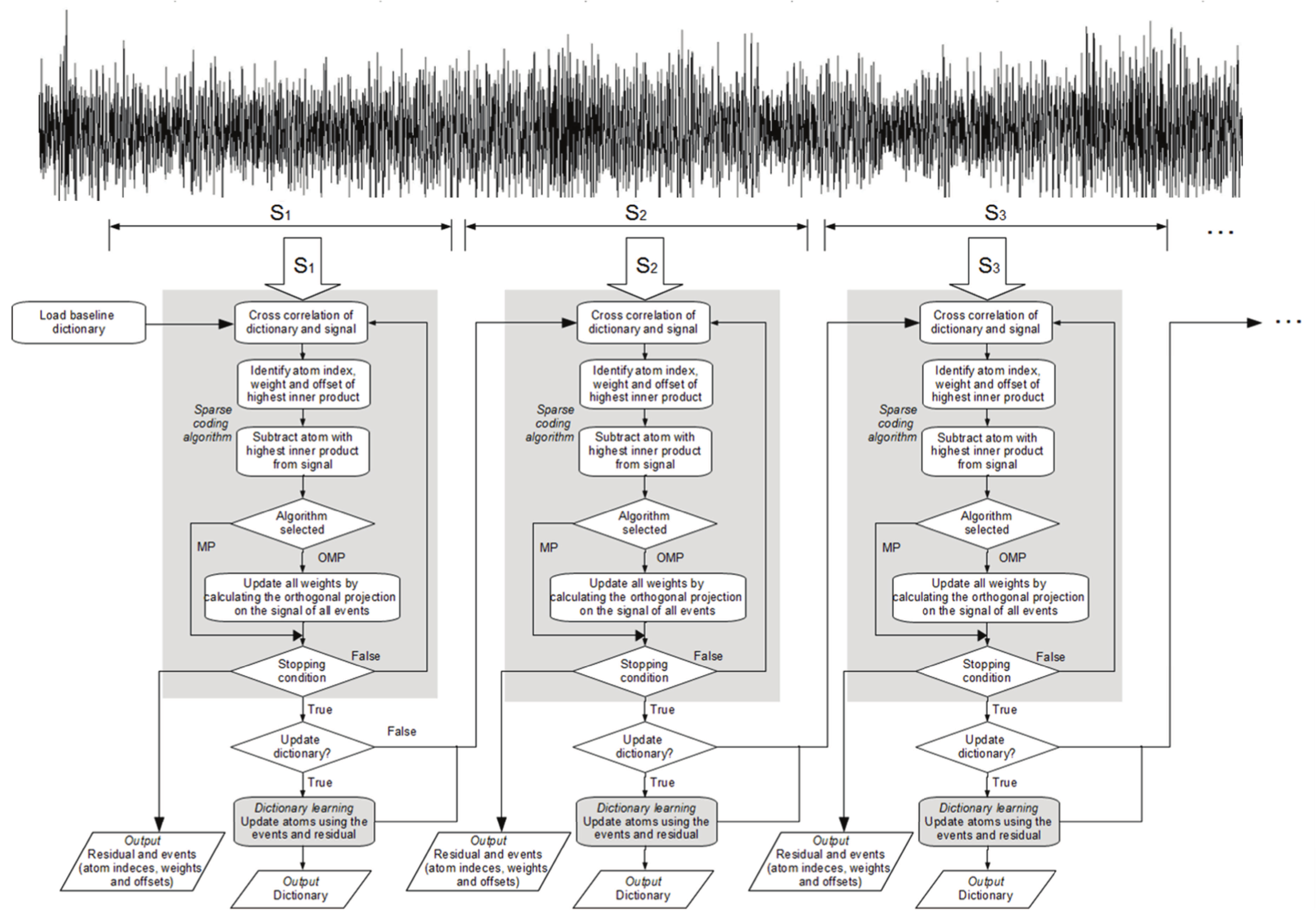}
\caption{
Dictionary learning scheme for online condition monitoring.
}
\label{fig:DLscheme}
\end{figure}

\fig{DLscheme} presents \blue{an} online monitoring scheme based on dictionary learning.
The signal is divided \blue{into} segments of equal length.
The interval between \blue{the} processed segments can be adapted to match \blue{either} the processing capacity of the condition monitoring system or the availability of \blue{the} data communicated from the turbine (which is the case considered here).
In the latter case, the interval between \blue{the} segments can be up to hours \blue{or} days due to \blue{the} limitations of the communication network between the wind farm and the condition monitoring center.
In an online processing scenario, \blue{the} edge effects due to signal segmentation can be reduced by transferring the tail of the residual to the next segment to be processed.
Martin-del-Campo et al.~\cite{MartindelCampo2016} describes this method for processing continuous signals.

The \red{initial dictionary is either pseudorandomly generated (training stage) or copied from a repository that includes dictionaries learned from similar machines (monitoring stage) according to a experimental protocol such as the one defined in \sect{methodology}.}
Initially, the first segment is processed with the sparse coding algorithm (MP or OMP), and the resulting sparse representation is used to update the dictionary.
Subsequently, the updated dictionary is used to process the next signal segment, which is a process \blue{known as} dictionary propagation.
The output of this process is the updated dictionary, the residual, and the coefficients and offsets of the selected atoms that define the sparse approximation of the signal.
These parameters are used as features for monitoring the corresponding wind turbine.

Note that dictionary learning can be deactivated by setting the learning rate parameter $\eta$ to zero.
In that case, the dictionary is constant over time.
However, the sparse representation of the signal can \blue{still} be generated and analyzed for condition monitoring purposes.

\subsection{Dictionary distance}

The dictionary is updated when each signal segment is processed (provided that the learning rate is nonzero).
Therefore, it is interesting to quantify and monitor the difference between two dictionaries learned at different points in time, for example, by comparing the present dictionary to a baseline dictionary learned during a period when the machine was known to be in a healthy condition.
Skretting and Engan \cite{Skretting2011} define the distance between two dictionaries $\Phi'$ and $\Phi$ as
\beq
\red{\beta (\Phi,\Phi') = \beta (\Phi',\Phi) = \frac{1}{2M} \Big(\sum_{i=1}^M \beta (\Phi',\phi_j) + \sum_{j=1}^{M} \beta (\Phi,\phi'_j) \Big),}
\label{eq:DictChange}
\eeq
\red{where the two dictionaries have the same number of atoms, $M$, with memberships $\phi\in\Phi$ and $\phi'\in\Phi'$.}
The function $\beta$ is the maximum similarity of an atom $\phi$ to the atoms in dictionary $\Phi$, 
\beq
\red{\beta (\Phi,\phi) = \arccos \mu(\Phi,\phi),}
\label{eq:AtomChange}
\eeq
where $\mu$ is the mutual coherence, which is defined \red{as follows~\cite{elad2010sparse}
(here, the function $\mu$ should be a function of both $\Phi$ and $\phi$ according to how it is used in~\cite{elad2010sparse}):}
\beq
\red{\mu ( \mathbf{\Phi}, \phi_i) = \max_{
		\substack{
			\forall\phi_j\in\Phi \\
			j\neq i
		}
	} \frac{| \mathbf{\phi}_i^T \mathbf{\phi}_j |}{\| \mathbf{\phi}_i \|_2 \cdot \| \mathbf{\phi}_j \|_2}.}
\label{eq:Coherence}
\eeq
The dictionary distance is measured in degrees and conceptually is a generalization of the conventional ``cosine of angle'' measure of similarity.
For example, when $\beta (\Phi,\Phi') = 0$ the two dictionaries are equal.

The dictionary distance measure can be used to quantify the distance between one learned dictionary at two different points in time.
We refer to this measure as the adaptation rate (of the dictionary) and define it as $\beta (\Phi^t,\Phi^{t-\delta})$, where $t$ is the current time and $\delta$ refers to the same dictionary at some point in the past.
The idea is that the adaptation rate can be used to quantify sudden abnormal changes in the signal, for example due to a fault in the system.
See also \cite{MartinDelCampo2015} where the rate of change of individual atoms after the introduction of a fault is investigated.
In principle, the dictionary distance $\beta (\Phi^t,\Phi^{t-\delta})$ could be normalized with the time step $\delta$ to obtain a finite difference approximation of the ``dictionary derivative'' with respect to time.


\section{Data source}
\label{sec:datasource}

\begin{figure}[h]
	\centering
	\includegraphics[width=0.75\textwidth]{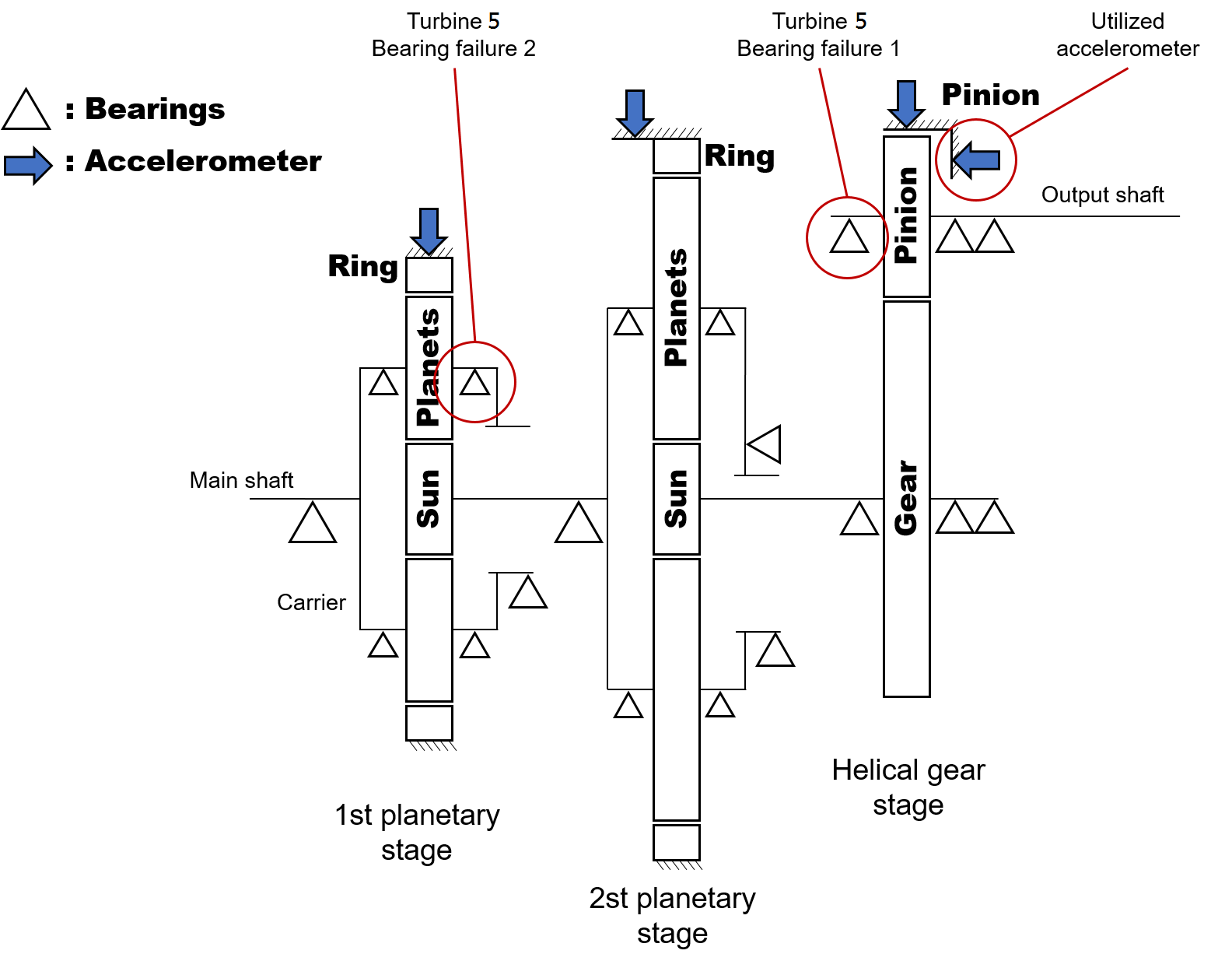},
	\caption{Schematic view of the gearbox in a wind turbine.
	 The components of each stage are shown, including the support bearings.
     Data from one wind turbine with two bearing failures \blue{are} included in this study.
     The locations of the faulty bearings are highlighted in the figure.
	 The measurement axes of the accelerometers are marked by arrows.
	 }
	\label{fig:Schematic}
\end{figure}

We aim to study the viability of a dictionary learning approach to condition monitoring using \blue{real--wolrd data. As a result} we have no control over the operational and environmental conditions, \blue{which is} in contrast to former studies based on data from controlled experiments.
The data originate from a wind farm located in northern Sweden.
The wind turbines are the same model and have integrated condition monitoring systems that transmit data to a condition monitoring database\blue{; from this database, we can access} the vibration data used in this study.
Each wind turbine \blue{posseses} a three-stage gearbox, including two sequential planetary gear stages followed by a helical gear stage.
Each gearbox has four accelerometers located near the different gear stages.
\fig{Schematic} includes a schematic view of the gearbox and the \blue{locations} of the accelerometers.

\begin{figure}[h]
	\centering\includegraphics[width=0.40\textwidth]{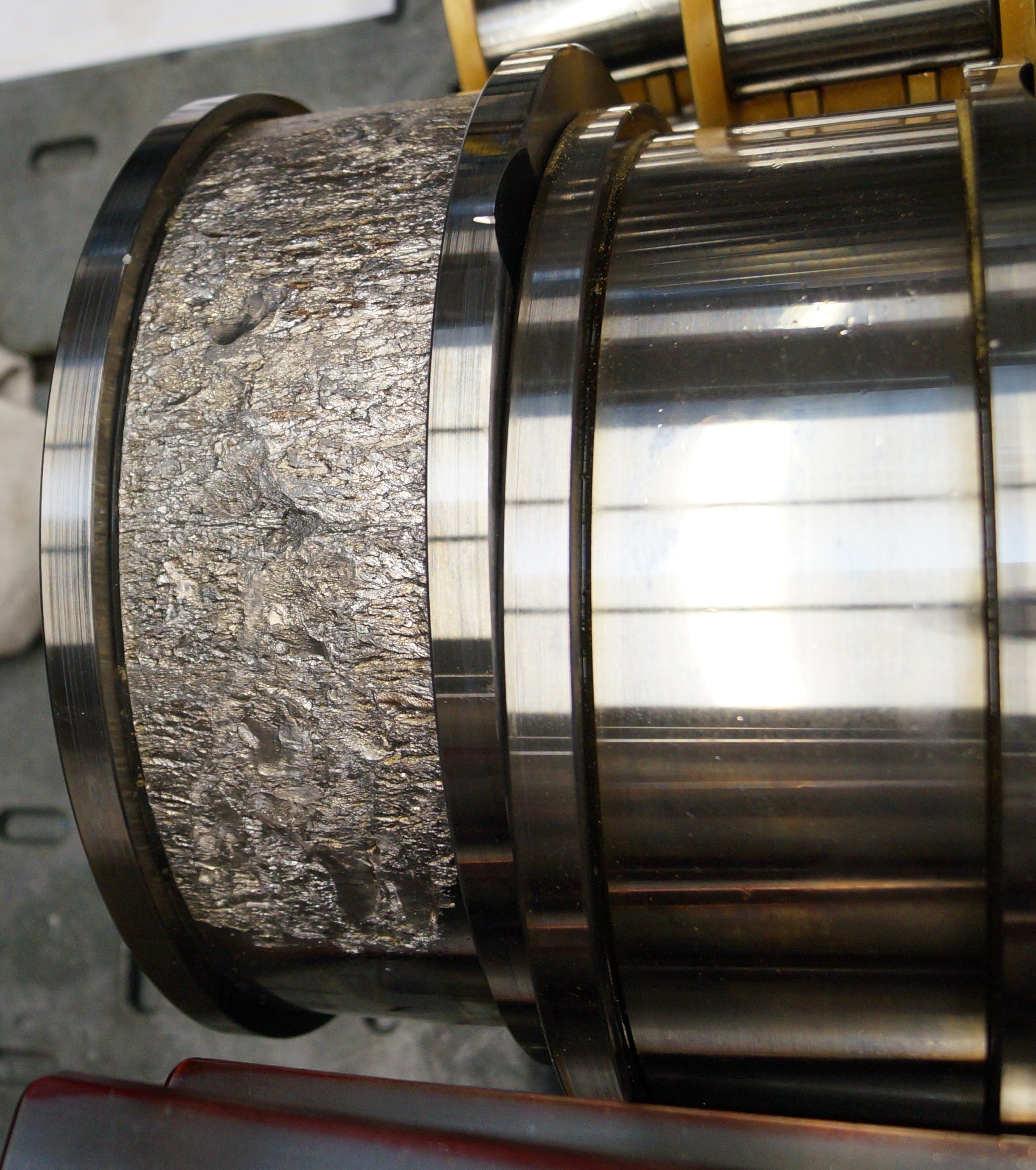}
	\caption{End result of the inner raceway failure of the bearing supporting the planets in Turbine 3 (left).
	Inner raceway of a healthy bearing included for comparison (right).
	}
    \label{fig:PlFailure}
\end{figure}

Raw time-domain vibration signals from six turbines within the same wind farm are considered in this study.
All \blue{of the} measurement data correspond to the axial direction of the accelerometer\blue{,
which} is mounted on the housing of the output shaft bearing of each turbine.
The sampling rate is 12.8~\red{kHz,} and each signal segment is 1.28 seconds long (16384 samples).
The signal segments \blue{were} recorded with an interval of approximately 12 hours over a period of 46 consecutive months in the last decade.
\blue{In} this period of time, five turbines \blue{remained healthy, which will henceforth be} referred to as Turbine 1, Turbine 2, Turbine 3, Turbine 4 and Turbine 6.
The \blue{other turbine, which we refer} to as Turbine 5, had two bearing failures \blue{in this period}.
The locations of the defective bearings are highlighted in \fig{Schematic}.
The descriptions of the failures are \blue{as follows}:
\begin{enumerate}
	\item Inner raceway failure on a four-point ball bearing on the output shaft.
	\blue{The output} shaft bearing was replaced after 1.2 years in operation.
	\item Inner raceway failure on one of the four cylindrical roller bearings supporting one of the planets in the first planetary gear.
	The entire gearbox was replaced after 2 years in operation.
	\fig{PlFailure} shows the end result of this failure.
\end{enumerate}
The dataset containing the raw time-domain vibration signals and speed measurements from the six turbines is publicly available; see \app{data} for further information.

\section{Results and discussion}
\label{sec:results}

\subsection{Outline of numerical experiments}
\label{sec:methodology}

In addition to investigating the bearing failures in Turbine~5 described above,
we are interested in the \blue{similarities of the} of dictionaries learned from different wind turbines of the same type that are located in the same geographical area.
Can the dictionary learned from one turbine be similar to the dictionary learned from \blue{a} similar turbine that is subject to similar operational and environmental conditions?
Furthermore, is a dictionary learned from one healthy turbine useful for monitoring of \blue{a} similar turbine?
To address these questions, we analyze the vibration signals described in \sect{datasource} using the dictionary learning method described in \sect{dictionary_learning_approach}.
We process the data with our \blue{MATLAB/C++} implementations of MP and local OMP~\cite{Mailhe2009} \blue{as well as} Smith and Lewicki's dictionary learning algorithm~\cite{Smith2006}.
There are two \red{main} stages in the experimental protocol \blue{we used:} \red{1) learning of a baseline dictionary for each turbine with data from a period of non-faulty operation (training stage) and 2) updates of each turbine dictionary using the successive recorded signal segments (monitoring stage).
For each turbine, the baseline dictionary defines the first dictionary used in the monitoring stage.}

\subsubsection{\red{Training stage}}

\begin{figure}[!h]
\centering
\includegraphics[width=0.95\linewidth]{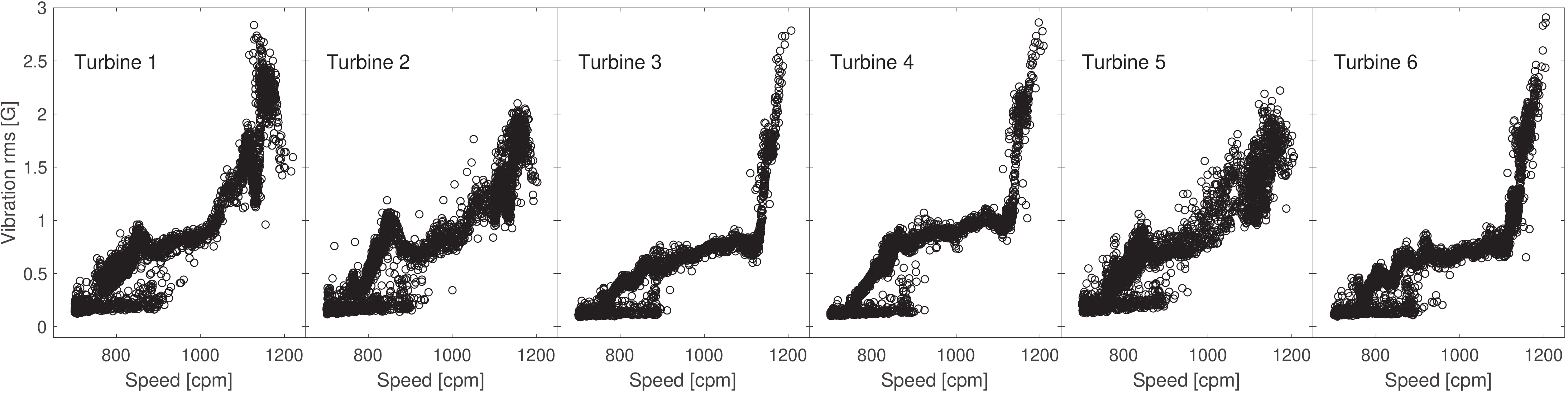}
\caption{
Scatter plot of vibration RMS of each recorded signal segment vs. the speed in cycles per minute (cpm) for the six turbines.
The data shown correspond to the full recording period \blue{of 46 months}.
}
\label{fig:VibRMS}
\end{figure}

\red{In the training stage the aim} is to learn a baseline dictionary for each turbine that corresponds to the signal recorded \blue{under healthy/non-faulty} operational conditions.
We use the signal segments recorded in the time period comprising approximately \blue{the period} after the second year of operation of the turbines.
\blue{Because we selected the} period of time after the replacement of the gearbox in Turbine~5, \blue{this turbine also operated} in a healthy condition during that time period.
We use the same training period for the six turbines to ensure similar operational and environmental conditions during training.
Signal segments with a vibration RMS above 0.5~G are included in the training process, while segments with a lower RMS are omitted from the analysis presented here.
\fig{VibRMS} shows the RMS values of the signal segments versus the rotational speed.
Below an RMS of 0.5~G, the turbines are sometimes unloaded and the corresponding signal segments are more noisy (possibly due to the reduced load applied to the bearings).
Thus, we introduce a threshold on the RMS to exclude signal segments recorded when the turbines/gearboxes are unloaded.
\tab{sel_sigseg} presents a summary of the signal segments available and considered in the training stage.
Training is performed with 5000 signal blocks \red{(used column) with a duration} of one second (12800 samples).
Each one-second block is randomly selected from within the signal segments \blue{with durations} of 1.28~seconds (16384 samples).
The signal segments are randomly selected from the time period comprising the third year of operations, \red{where the total number of available segments is shown in the column ``available''.
The considered column lists the number of segments after discarding segments with a vibration RMS below 0.5~G.
Each block is preprocessed} to have zero mean and unit variance.
Both sparse coding algorithms are stopped at 90\% sparsity, which means that each block of 12800 samples \blue{is} modeled with 1280 atom instances.
We use a step size of $\eta = 10^{-6}$ for the dictionary update.

Before the first signal block \red{in the training stage} is processed, the dictionary is initialized with a pseudorandom dictionary of eight atoms.
Initially, the atoms are seventy elements long and are always defined in the same way at the start of the learning process.
Each atom is generated from fifty elements sampled from a Gaussian distribution with zero mean \blue{that} is zero-padded with ten samples at each tail.
The atoms can grow in length during the learning process\blue{, and they} are normalized after each update.
The dictionary learned after the first 5000 iterations is henceforth referred to as a baseline dictionary, and it is used as \blue{the} initial dictionary in the \red{monitoring stage}.
This procedure is repeated for the six turbine cases \red{at the beginning of their training stage using the same pseudorandom dictionary.
Consequently,} a baseline dictionary is generated for each turbine.

\begin{table}[!h]
                \captionsetup{singlelinecheck=off,margin=0cm}
        \caption{Number of available and selected signal segments.}
\small
\vspace{-0.6cm}
\begin{center}
	\begin{tabular*}{1.00\textwidth}{@{\extracolsep{\fill}} lccccc}  
	\toprule
	& \multicolumn{3}{c}{Training} & \multicolumn{2}{c}{\red{Monitoring}} \\
	\cmidrule(r){2-4}  \cmidrule(r){5-6}
	Case       & Available & Considered & Used*   & Available & Used \\
	\midrule
	Turbine 1  & 1212      & 859        & 5000   & 2982      & 2078    \\
	Turbine 2  & 1203      & 810        & 5000   & 3005      & 2058    \\
	Turbine 3  & 1243      & 527        & 5000   & 2670      & 1135    \\
	Turbine 4  & 1248      & 768        & 5000   & 2667      & 1623    \\
	Turbine 5  & 1237      & 803        & 5000   & 2976      & 1907    \\
	Turbine 6  & 1220      & 642        & 5000   & 2953      & 1629    \\
	\bottomrule
	\end{tabular*}
	\raggedright * The processed blocks are one second long and are sampled at random offsets in the signal segments.
\end{center}
        \label{tab:sel_sigseg}
\end{table}

\subsubsection{\red{Monitoring stage}}

The \red{monitoring stage is similar for the six turbine cases and is representative of the online monitoring scheme described in \fig{DLscheme}.
The learned dictionaries in the training stage, which are now known as baseline dictionaries, are used as input into the sparse coding with dictionary learning algorithm.}
In this second processing stage, we consider all \blue{of the available signal segments} during the 46 consecutive months of data.
However, \blue{as} in the first learning stage, we use only segments with a vibration RMS value above 0.5~G.
\tab{sel_sigseg} includes a summary of the number of used signal segments for each turbine.
\red{The available column lists all of the available segments for each turbine, and the used column lists the number of segments after discarding those with a vibration RMS below 0.5~G.} 
Each segment is preprocessed to have zero mean and unit variance.
\blue{As} before, the MP and OMP algorithms are stopped at 90\% sparsity, which means 1600 atom instances are used to model 16384 samples.
The signal segments are analyzed in sequential order, as would be the case in an online monitoring situation.
Therefore, the dictionary is said to be propagated over time, which means that it is gradually adapting to the structure of the signal.
A method for edge effect reduction is not introduced in this stage because the learned translation-invariant atoms are about two orders of magnitude shorter than the processed signal window and the existing 12-hour gap between signal windows. 

Two scenarios are considered in the \red{monitoring} stage.
In the first scenario, the dictionary is propagated with a step size of $\eta = 10^{-6}$.
In this scenario, we study the change \blue{in} the propagated dictionary over time with respect to the baseline dictionary.
In the second scenario\blue{, the dictionary remains} constant by setting $\eta = 0$.
In this scenario, we study the fidelity of the sparse representation model over time using the fixed baseline dictionary.
In the second scenario, we are interested \blue{in studying} the effect of the bearing failures on the fidelity of the sparse model.

The \red{monitoring} stage continues with \blue{the} testing of two additional cases.
\blue{These cases} focus on the importance of the baseline dictionary by investigating the consequences of propagating a dictionary that is not optimized to the signal of the machine.
Signal segmentation and \blue{preprocessing is performed} in the same way as in the previous two \red{monitoring} cases.
In one case, we use a baseline dictionary learned from each turbine to model and analyze the signals from the remaining five turbines.
We repeat this procedure with the six turbines.
In the other case, we use an arbitrary baseline dictionary learned from vibration signals obtained from the ball bearing data center at Case Western Reserve University~\cite{Loparo2003}.
In the latter case the signals are generated by a rotating machine consisting of an electric motor, a torque transducer, a dynamometer and a ball bearing supporting the motor shaft.
An accelerometer located at the drive end of the motor is used to record the vibration data with a sampling rate of 12~\red{kHz}.
We alternate between several recorded datasets from a healthy bearing to simulate a varying load between 0 HP and 3 HP.
Thus, the dictionary used in this case does not encode information about the wind turbine signals and \blue{is} not expected to result in particularly accurate sparse codes of the vibration signals.

\subsection{Generalization across turbines}

\begin{figure}[!h]
	\centering
	\begin{subfigure}{.5\textwidth}
		\centering
		\includegraphics[width=.95\linewidth]{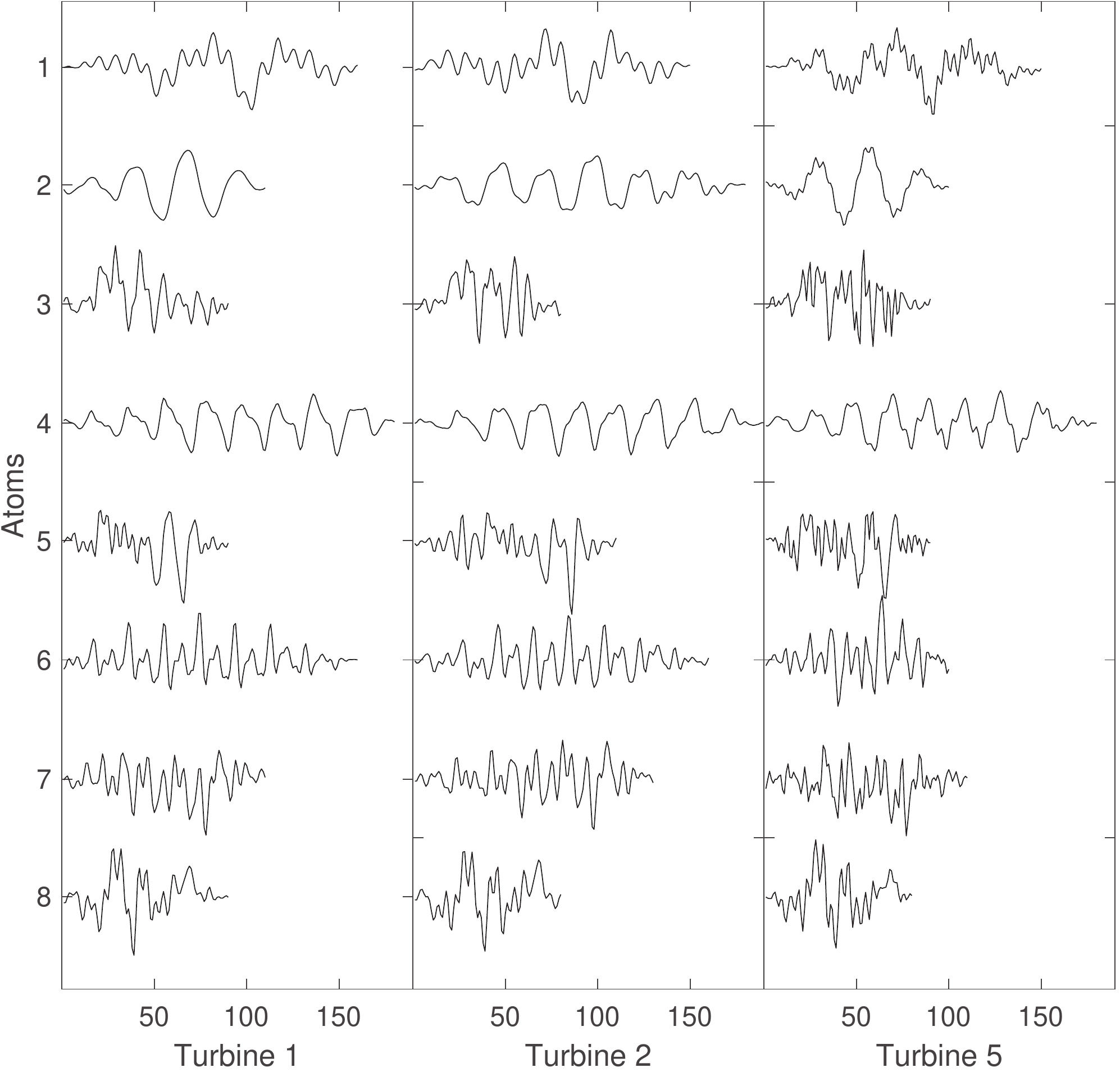}
		\caption{MP}
		\label{fig:TrainDictMP}
	\end{subfigure}%
	\begin{subfigure}{.5\textwidth}
		\centering
		\includegraphics[width=.95\linewidth]{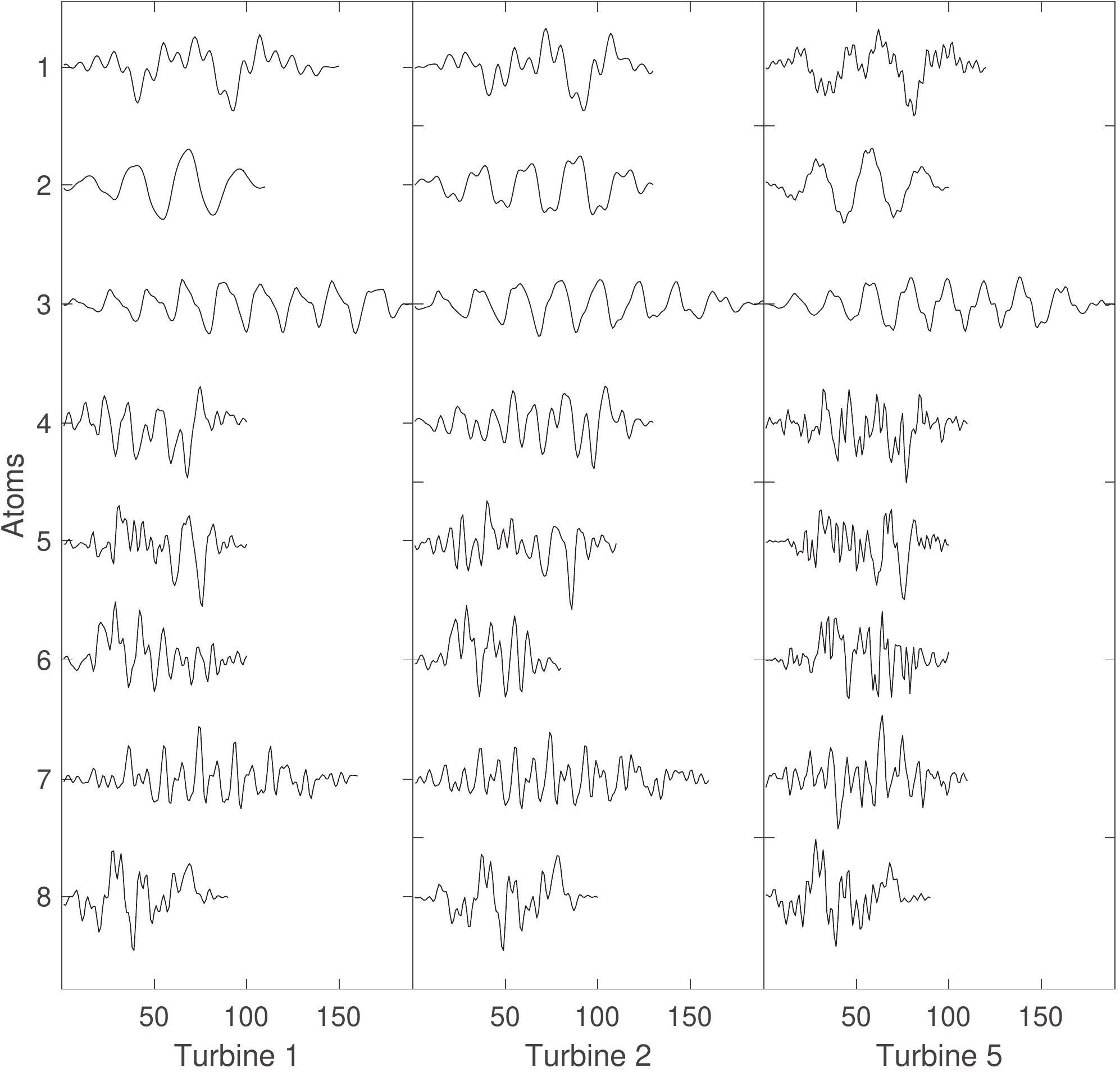}
		\caption{OMP}
		\label{fig:TrainDictOMP}
	\end{subfigure}
	\caption{
		Atoms learned from vibration signals from \blue{the} selected three turbines \red{at the end of the training stage}.
		Turbine 1 did not have \blue{any} reported failures. 
		Turbine 2 had an electrical sensor fault at the beginning of its operation.
		Turbine 5 had the faults described in \sect{datasource}.
		The atoms of Turbine 1 are ordered by \blue{the} ascending center frequency.
		The atoms of Turbine~2 and Turbine~5 are ordered by maximizing the cross-correlation with respect to the atoms of Turbine 1.
		All atoms are normalized.
	}
	\label{fig:TrainDict}
\end{figure}

In the training stage, one baseline dictionary is learned for each turbine \blue{under} healthy conditions.
Here, we aim to investigate how similar the learned dictionaries are across different turbines.
Thus, we process the signal from the same accelerometer location in the six turbines.
Furthermore, we are interested in \blue{the differences that result} from the use of two different sparse coding algorithms, \blue{namely,} MP and OMP.
We use the same protocol and hyperparameters during learning with MP and OMP.
The resulting dictionaries for three turbines using both algorithms are shown in \fig{TrainDict}, \red{and each dictionary includes eight updated atoms.}
The selected dictionaries include the following known cases: a healthy condition for Turbine 1, an electrical sensor failure for Turbine 2 and a gearbox failure for Turbine 5.
However, all \blue{of the} baseline dictionaries are trained during a period of time when the turbines are expected to be healthy (meaning that no faults were detected during or after that period of time).  
The dictionaries are obtained after learning from 5000 signal blocks,
which \blue{correspond to approximately} 83~minutes of vibration data and 64~million samples.
The \red{updated atoms have different length and (L2) normalized magnitudes, and all atoms are illustrated in the same scale in the two} panels.
The atoms corresponding to Turbine~1 are ordered by \blue{the} ascending center frequency in both sparse coding algorithm cases.
The atoms of Turbine~2 and Turbine~3 are ordered in the corresponding way by maximizing the cross-correlation with each atom of Turbine~1.
\tab{centerfreq} summarizes the center frequencies of the atoms.

\begin{table}[!h]
	\captionsetup{singlelinecheck=off,margin=0cm}
	\caption{Center frequencies of \red{atoms in the baseline dictionary} for MP (OMP).}
	\small
	\vspace{-0.6cm}
	\begin{center}
		\begin{tabular*}{1.00\textwidth}{@{\extracolsep{\fill}} lcccccccc}  
			\toprule
			& \multicolumn{8}{c}{Center frequency [kHz]} \\
			\cmidrule(r){2-9}
			Case       & 1 & 2 & 3 & 4 & 5 & 6 & 7 & 8 \\
			\midrule
			Turbine 1  & 0.28(0.26) & 0.47(0.47) & 0.64(0.64) & 0.64(0.70) & 0.92(0.90) & 1.36(1.02) & 1.40(1.36) & 1.49(1.49)  \\
			Turbine 2  & 0.50(0.29) & 0.61(0.49) & 0.86(0.60) & 0.88(1.28) & 0.99(1.45) & 1.36(1.04) & 1.52(1.36) & 1.97(0.70)  \\
			Turbine 3  & 0.14(0.36) & 0.56(0.56) & 0.76(0.64) & 0.96(0.76) & 1.00(1.01) & 1.11(1.11) & 1.28(1.23) & 1.60(1.60)  \\
			Turbine 4  & 0.28(0.57) & 0.80(0.75) & 0.92(0.80) & 0.99(0.93) & 1.01(1.05) & 1.28(1.22) & 1.28(1.46) & 1.45(1.66) \\
			Turbine 5  & 0.26(0.32) & 0.45(0.45) & 0.64(0.61) & 0.92(1.80) & 1.60(0.96) & 1.60(1.86) & 1.85(1.69) & 2.68(1.56)  \\
			Turbine 6  & 0.57(0.32) & 0.96(0.60) & 0.96(0.69) & 1.02(1.09) & 1.23(1.28) & 1.32(1.28) & 1.60(1.66) & 2.40(2.38)  \\
			\bottomrule
		\end{tabular*}
	\end{center}
	\label{tab:centerfreq}
\end{table}

Some of the atoms learned from the three turbines appear \blue{to be} similar, while a few are different, as shown in \fig{TrainDict}.
Furthermore, there are similarities between the atoms learned with the MP and OMP methods.
For example, atoms one and two have sinusoidal components of relatively low frequency in all three cases regardless of the sparse coding method used.
Atoms three and four are exchanged between the MP and OMP cases.
Atom four with MP and atom three with OMP have a clearly visible sinusoidal component.
In contrast, atom three with MP \blue{has} a smaller central frequency and \blue{a} more noise-like appearance compared to atom four with OMP.
Atoms five to eight are more noise-like \blue{and have} higher center frequencies in both cases.
Atoms six and seven are the most different across the three turbines when using MP or OMP.
However, though they are different across the turbines, atoms six and seven are similar for MP and OMP, in particular, for Turbine~1 and Turbine~2 but not for Turbine~5.

Note that \blue{thought all of} the dictionaries are trained using healthy signal segments recorded under similar operational and environmental conditions, the bearings are not identical.
The bearings in Turbine~5 are newer compared to the bearings in Turbine~1 and Turbine~2 due to the preceding gearbox replacement.
The bearings in Turbine~1 and Turbine~2 have been in operation for more than two years.
Thus, these bearings have been degraded compared to the new bearings in Turbine~5.
Therefore, we cannot expect the \red{updated atoms} \blue{to} be identical for the three turbines illustrated here.

\subsection{Effect of baseline dictionary selection}

\begin{figure}[!b]
	\centering
	\begin{subfigure}{.9\textwidth}
		\centering
		\includegraphics[width=.95\linewidth]{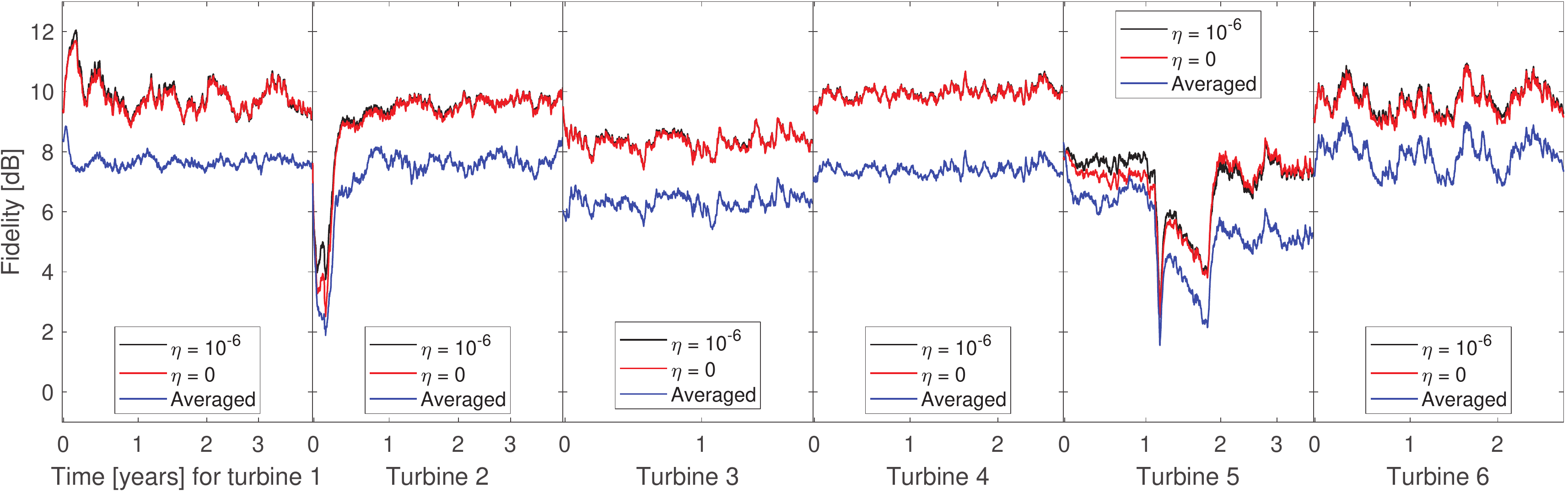}
		\caption{MP}
		\label{fig:FidWDMP}
	\end{subfigure}%

	\begin{subfigure}{.9\textwidth}
		\centering
		\includegraphics[width=.95\linewidth]{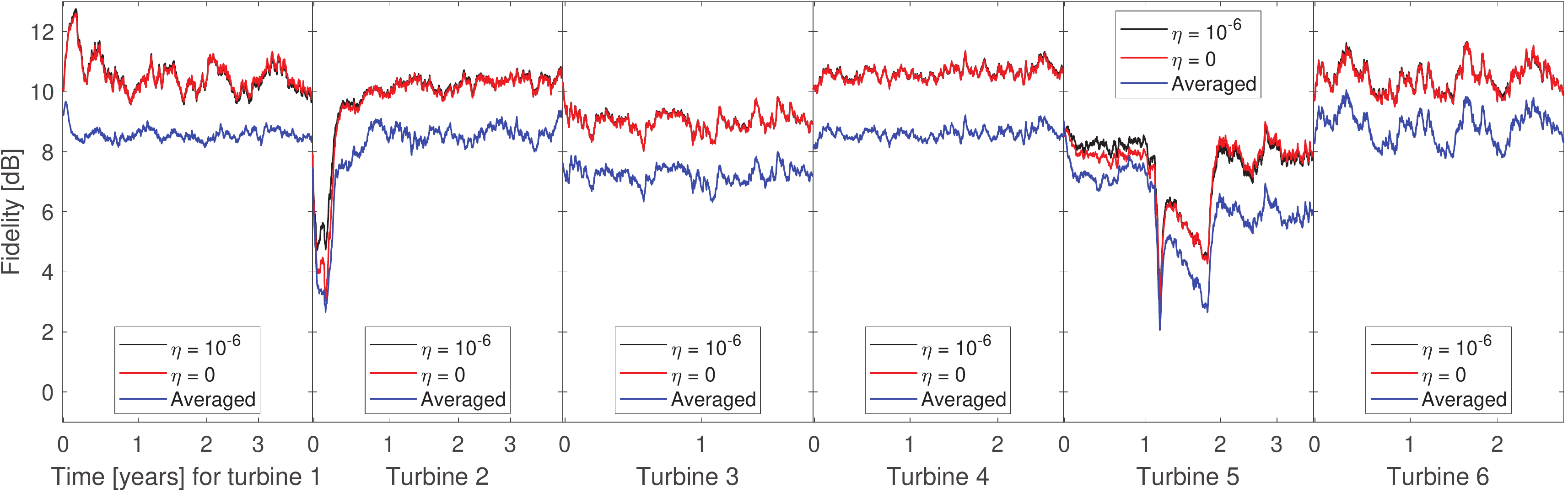}
		\caption{OMP}
		\label{fig:FidWDOMP}
	\end{subfigure}
	\caption{
		\red{Fidelity of the vibration signal model for $t \geq 0$ using MP (a) and OMP (b) for each turbine based on different dictionaries.
			A baseline dictionary is propagated over time in the $\eta = 10^{-6}$ case, while in the case of $\eta = 0$ the baseline dictionary is used as is, without further adaptation.
			The averaged case corresponds to the fidelity \blue{we obtain} when the baseline dictionary of each turbine is used without further modification to model the vibration signals of the other five turbines.}
	}
	\label{fig:FidWD}
\end{figure}

Next, we investigate \blue{the} effects of selecting different initial dictionaries and dictionary learning step lengths.  
In a field implementation of dictionary learning, the baseline dictionary needs to be learned from the signal to be monitored \blue{to ensure} that the model \blue{will have} high fidelity and effectively separate the signal from \blue{the} noise.
The fidelity of the model in \blue{decibels} is the ratio between the sparse approximation and the signal residual:
\beq
	dB = 20 log_{10} \left(  \frac{ \hat{s_{k}} }{\epsilon(t)} \right),
	\label{eq:FiddB}
\eeq
where $dB$ is \blue{the} fidelity in decibels, $\hat{s_{k}}$ is the sparse approximation of a signal segment and $\epsilon(t)$ is the corresponding residual.
\fig{FidWD} shows the model fidelities for three different cases.
In the first case, \red{which corresponds to $\eta = 10^{-6}$, the dictionary at $t=0$ is the baseline dictionary of each turbine; this dictionary} is propagated over time with a finite dictionary update step length.
In the second case, the dictionary update \red{is set at $\eta = 0$,} and the baseline dictionary is used as is without further modification.
In the third case, \red{which is labeled \textit{averaged}, the dictionary learned during the training stage of} each individual turbine is used to model the signals of the remaining five turbines without further modification of the dictionary.
\blue{For this case,} the average fidelity of the resulting five models is shown.
The fidelity is low-pass filtered with a first-order filter and a time constant of 15 days (30 signal segments) to improve the clarity of the plot.
The \red{$\eta = 10^{-6}$ and $\eta = 0$} cases result in similar fidelities, which indicates that the baseline dictionaries have converged and are not updated significantly by further learning.
When \blue{we use} the \red{averaged} baseline dictionary to model the signals from the other five turbines, the fidelities are slightly lower \blue{than when we use} the correct baseline dictionaries.
These results are observed independently of the sparse coding algorithm \blue{we} used.
MP has a slightly lower fidelity than OMP in the three cases considered. 
The large decrease \blue{in} the model fidelity for Turbine~5 in the beginning of year one corresponds to the period of time when the bearing was defective, and the two minima correspond to the replacement of the HSS bearing and the gearbox.
The low model fidelity for Turbine~2 at the beginning of year zero does not correspond to a previously known fault\blue{; this result is likely} due to an electrical issue after the installation of the sensor, cables or monitoring unit.

\begin{figure}[!h]
	\centering
	\begin{subfigure}{.95\textwidth}
		\centering
		\includegraphics[width=.95\linewidth]{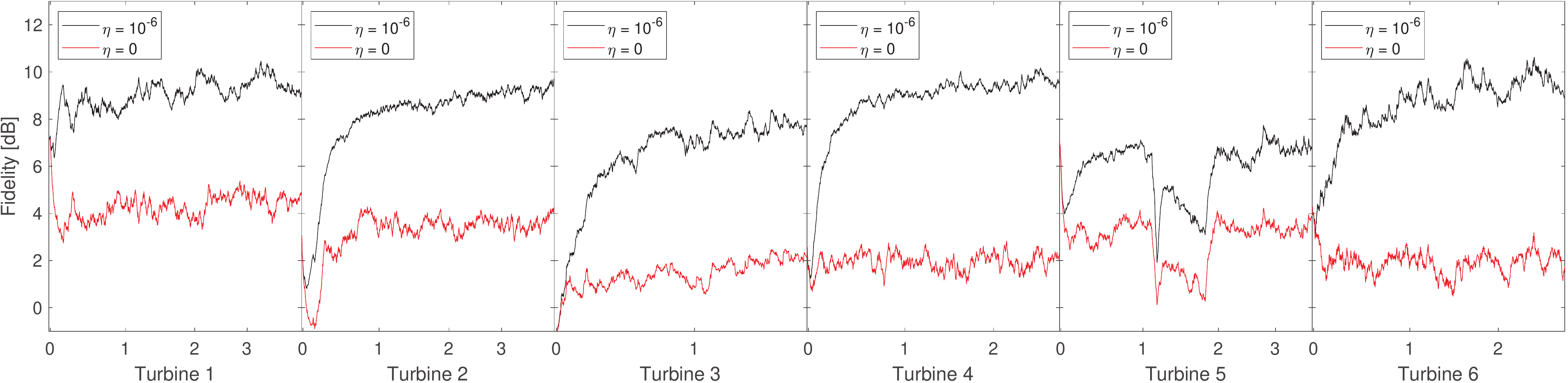}
		\caption{MP}
		\label{fig:FidBDCMP}
	\end{subfigure}%

	\begin{subfigure}{.95\textwidth}
		\centering
		\includegraphics[width=.95\linewidth]{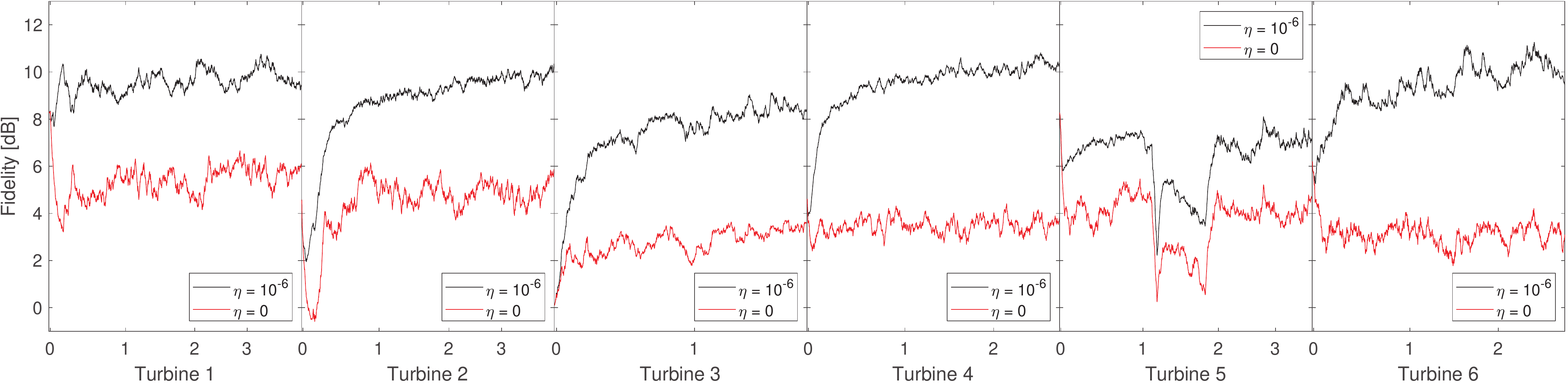}
		\caption{OMP}
		\label{fig:FidBDCOMP}
	\end{subfigure}
	\caption{
		\red{Fidelity of the signal models based on \blue{the} initial dictionaries learned from the CWRU database for $t \geq 0$ using MP (a) and OMP (b).
			In the case of $\eta = 10^{-6}$, the initial CWRU dictionary is propagated and adapted to the wind turbine vibration signal.
			In the case of $\eta = 0$, the baseline dictionary learned from the CWRU database is used as is.}
	}
	\label{fig:FidBDC}
\end{figure}

Next, two additional cases are considered where the \red{baseline} dictionary is learned from another machine.
In these cases, the vibration data \blue{were} obtained from the ball bearing data center~\cite{Loparo2003} at Case Western Reserve University (CWRU).
The fidelity of these two cases using the \red{ MP and OMP algorithms is shown in \fig{FidBDC}.
Each algorithm has two cases: $\eta = 10^{-6}$,} indicating when the \red{baseline} dictionary is propagated, and \red{$\eta = 0$, which occurs} when the \red{baseline} dictionary is used as is \blue{and} without further adaptation.
\red{Under the two algorithms, the fidelity in the $\eta =  10^{-6}$} case increases with time and reaches a similar level as in \fig{FidWD} at the end of the time period, which is to be expected since the atoms in the dictionary are adapted to the signal.
The CWRU baseline dictionaries cannot adequately model the wind turbine signals,
resulting in a lower fidelity in the \red{case of $\eta = 0$} compared to the fidelities in \fig{FidWD}.
However, note that the replacement of the output shaft bearing and gearbox in Turbine~5 are still associated with \blue{abnormally} low values of the fidelity.


\subsection{Distance to \blue{the} baseline dictionary}

The dictionary distance defined by \eqn{DictChange} quantifies the difference between two dictionaries.
This distance can be used to detect a gradual change \blue{in} a propagated dictionary by determining the distance between the propagated dictionary and \blue{either} the baseline dictionary or a set of baseline dictionaries.
Thus, faults that appear after a long period of degradation \blue{could} possibly be detected by monitoring the distance between the propagated dictionary and the baseline.
\blue{In principle, the} baseline dictionary is defined by atoms learned in normal states of operation; see \sect{methodology} for further details.

\begin{figure}[!h]
	\centering
	\begin{subfigure}{.5\textwidth}
		\centering
		\includegraphics[width=.9\linewidth]{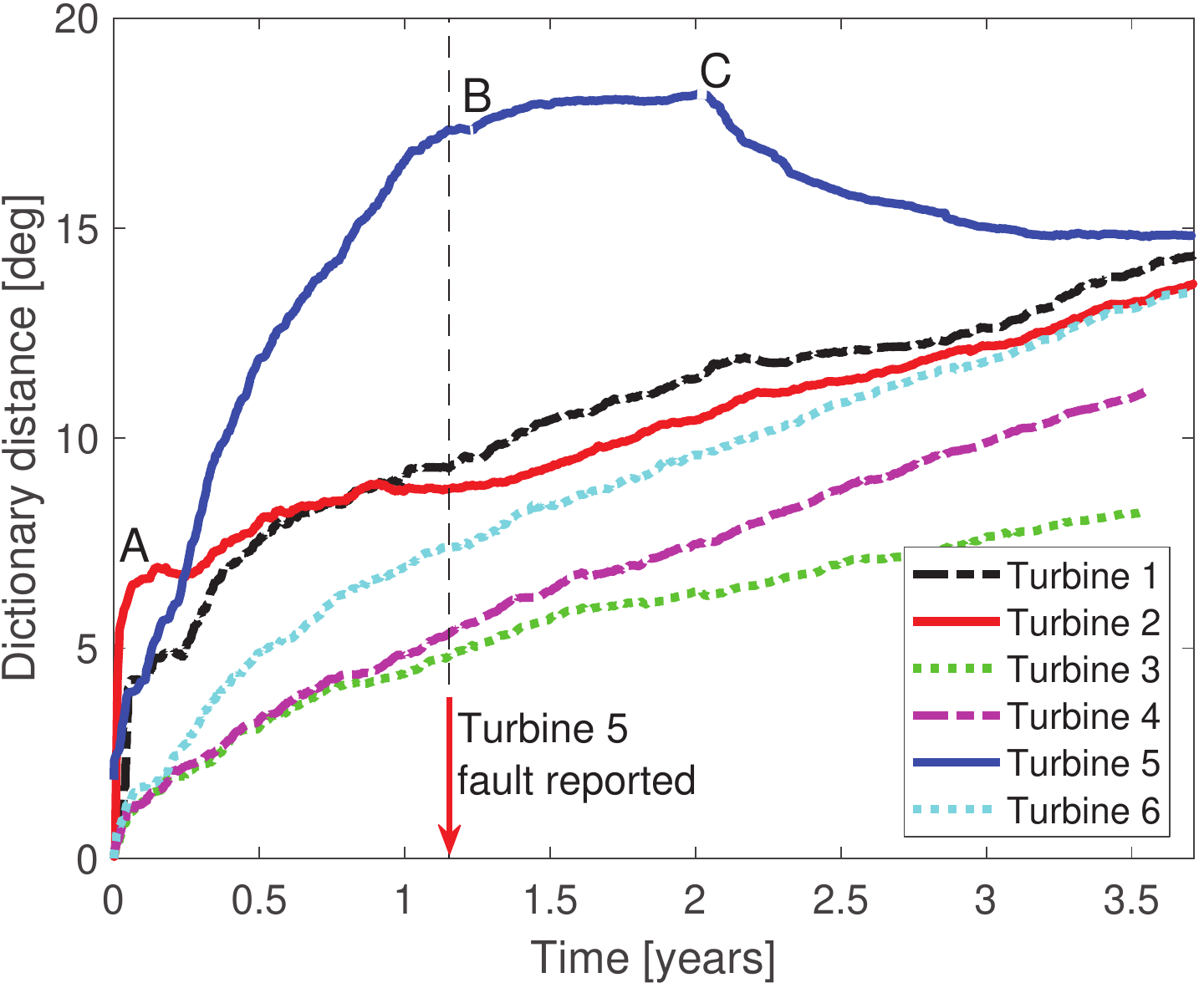}
		\caption{MP}
		\label{fig:Dict-distMPWD}
	\end{subfigure}%
	\begin{subfigure}{.5\textwidth}
	\centering
	\includegraphics[width=.9\linewidth]{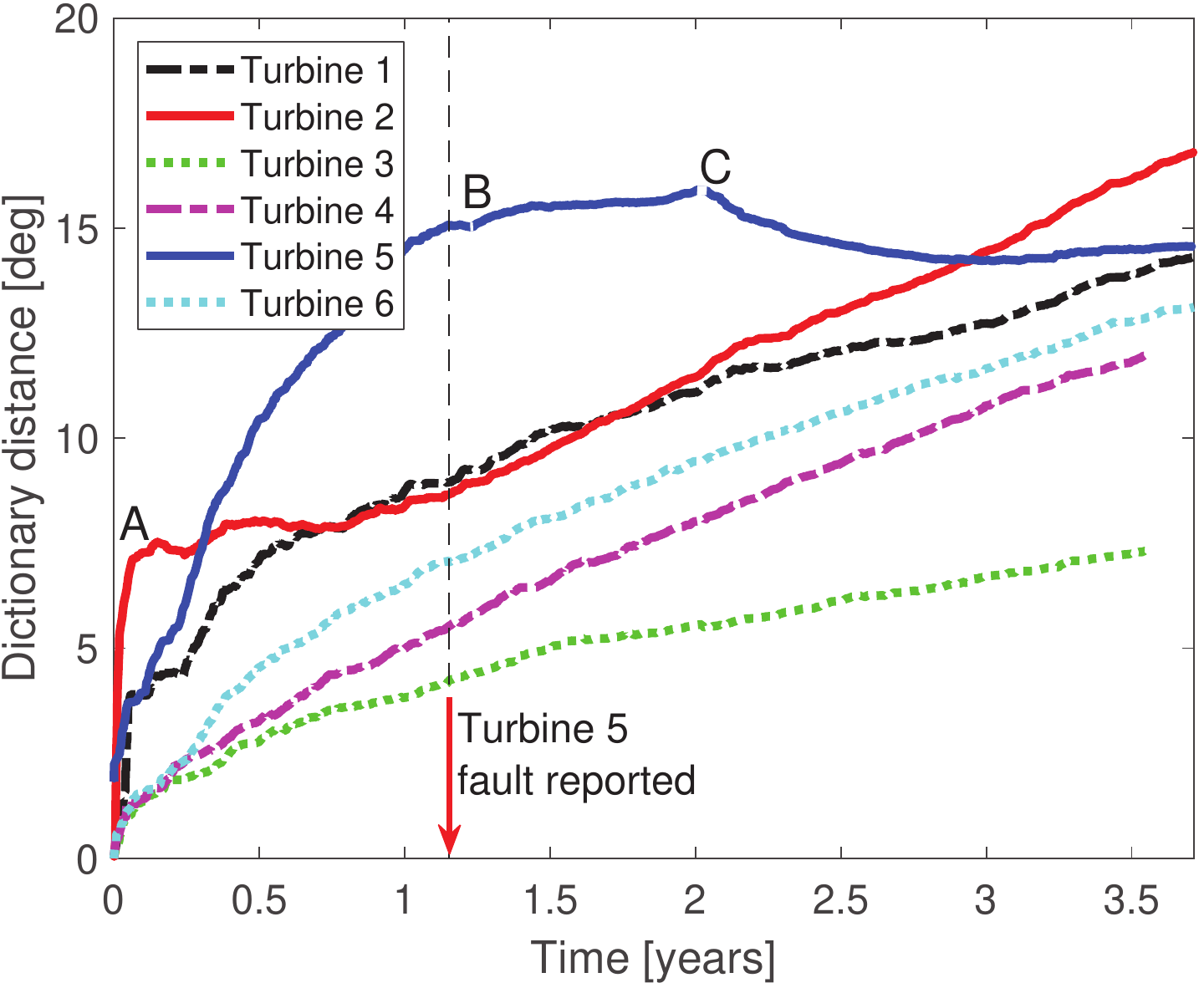}
	\caption{OMP}
	\label{fig:Dict-distOMPWD}
\end{subfigure}%
	\caption{
		Distance of a propagated dictionary with respect to the baseline dictionary versus time using the MP (a) and OMP (b) algorithms for the six turbines.
		\blue{Label} A indicates the end of a time period with a possible electrical fault in the data acquisition system of Turbine~2.
		\blue{In addition,} label B indicates the time period when the output shaft bearing was replaced in Turbine~5, and \blue{label} C indicates the subsequent gearbox replacement.
	}
	\label{fig:Dict-distboth}
\end{figure}

\fig{Dict-distboth} shows the corresponding dictionary distances for the six turbines using the two sparse coding algorithms.
As described above, the baseline dictionaries are trained with signal blocks recorded when the turbines are operating in healthy conditions (after the gearbox replacement in Turbine~5).
The resulting curve trends for both sparse coding algorithms describe similar behavior.
Turbine~1, Turbine~3, Turbine~4 and Turbine~6 show an increase in the dictionary distance when the dictionary is propagated over time.
For Turbine~2, there is a relatively fast initial increase \blue{in} the distance, which after some time stabilizes and becomes similar to the distance for Turbine~1.
The \blue{rapid} increase at the start is in agreement with the results presented in \fig{FidWD}\blue{; this figure} shows that the model fidelity is \blue{initially low, which is} most likely due to an electrical fault in the measurement system connected to the accelerometer.
In contrast, the dictionary distance for Turbine~5 increases \blue{more quickly} than the distances \blue{we} determined for all \blue{of} the other turbines.
The dictionary distance for Turbine~5 is \blue{approximately} two to three times higher than the distances for the other turbines at the point in time when the output shaft bearing is replaced in the gearbox.
After the bearing replacement, the distance is approximately stationary until \blue{it reaches} another peak just before the gearbox is replaced.
After \blue{the} replacement of the gearbox, the dictionary distance decreases and approaches the distance for Turbine~1, \blue{which indicates} the return to a normal condition.
The dictionary distances calculated with the OMP algorithm have a larger spread than the distances calculated with the MP algorithm.
\blue{While the} dictionary distance with the MP algorithm covers a 7-degree spread at the end of \blue{the recording period,} the dictionary distance with the OMP algorithm covers a 10-degree spread.

\begin{figure}[!h]
	\centering
	\begin{subfigure}{.48\textwidth}
		\centering
		\includegraphics[width=.9\linewidth]{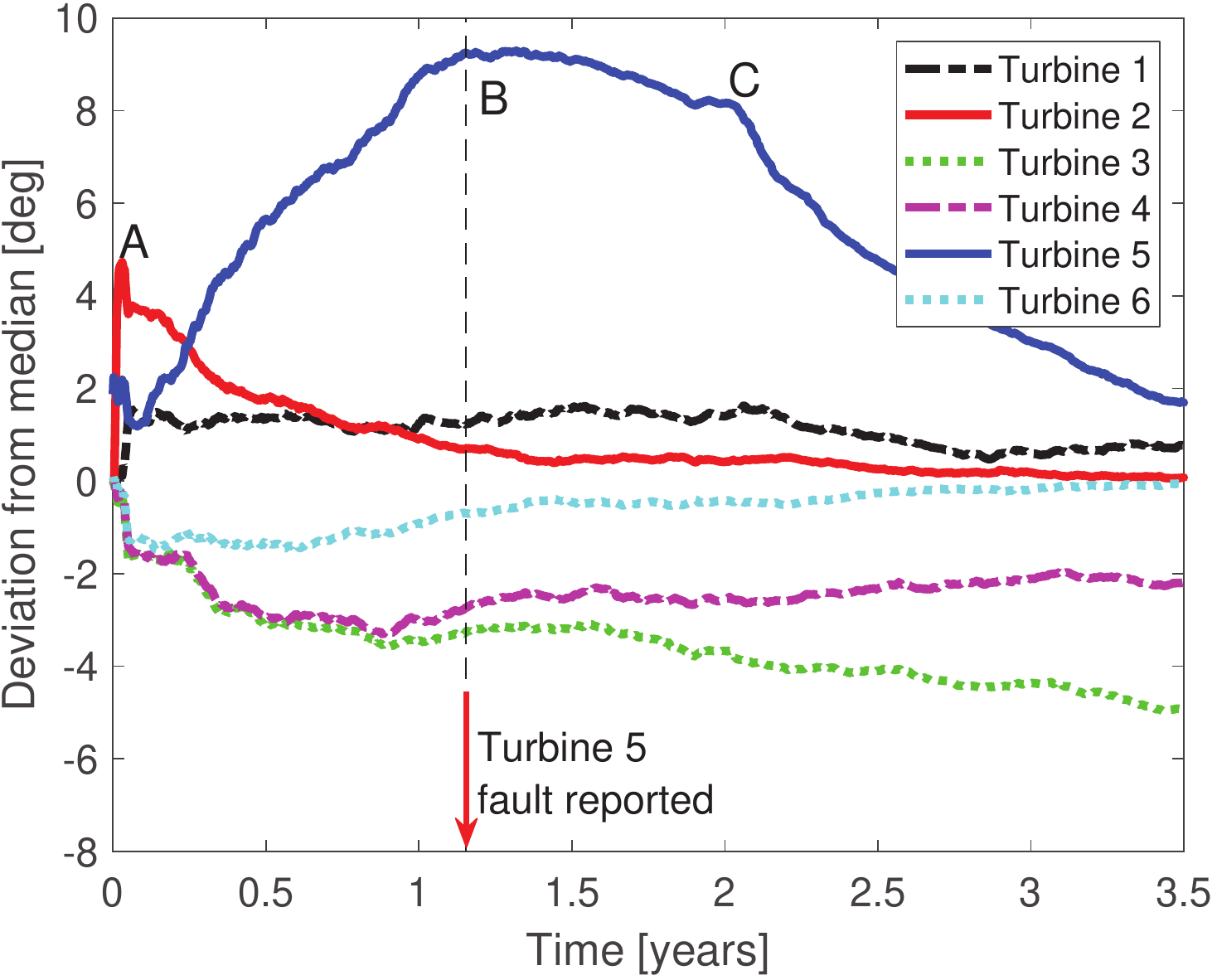}
		\caption{MP}
		\label{fig:madMPWD}
	\end{subfigure}
	\begin{subfigure}{.48\textwidth}
	  \centering
	  \includegraphics[width=.9\linewidth]{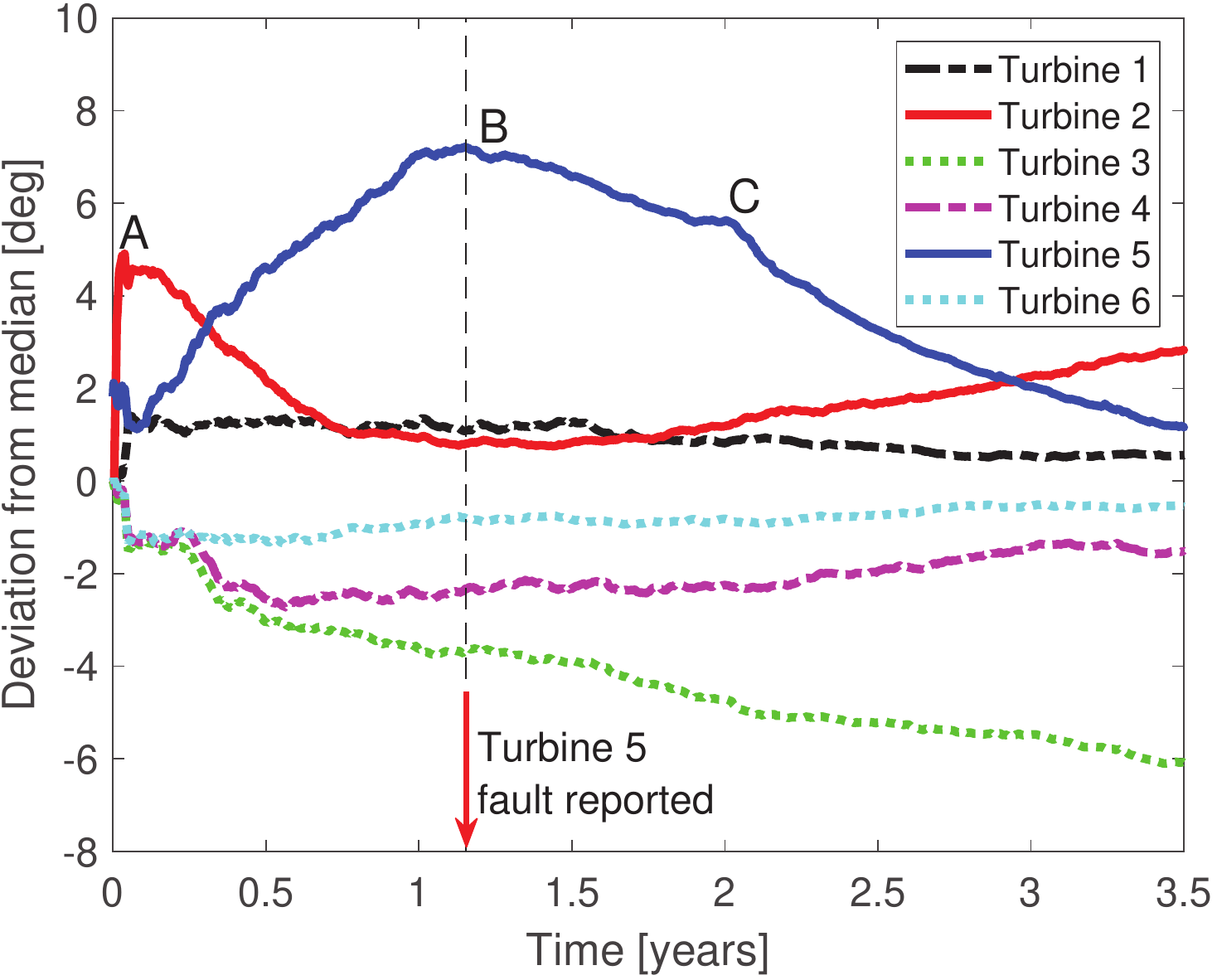}
	  \caption{OMP}
	  \label{fig:madOMPWD}
	\end{subfigure}
	\caption{
		Median absolute deviation of the dictionary distance versus time based on \blue{the} MP (a) and OMP (b) algorithms.
		\blue{Label} A indicates the end of a time period with a possible electrical fault in the data acquisition system of Turbine~2.
		\blue{In addition,} label B indicates the time period when the output shaft bearing was replaced in Turbine~5, and \blue{label} C indicates the subsequent gearbox replacement.
}
\label{fig:madboth}
\end{figure}

\blue{The dictionary} distance is an indicator that can be used for outlier detection in a population of monitored wind turbines.
\fig{madboth} shows the median absolute deviation (MAD) of the dictionary distance shown in \fig{Dict-distboth}.
The MAD highlights a significant deviation of Turbine~5 in a period that extends \blue{approximately} eight months earlier than the date when a fault report was filed for the turbine.
The fault report for Turbine~5 was filed 1.2 years after the start of \blue{the} vibration data recording.
The trend of Turbine~3 deviates from the other turbines at the end of the recorded time period for both sparse coding algorithms.
This deviation \blue{may be} due to the geographical location of Turbine~3, which is \blue{the} farthest away from the other turbines.

An adequate selection of the initial dictionary \red{under an online monitoring scheme is important to constrain the dictionary distance values and the} rate of change.
Using the baseline dictionaries trained on signals from the turbines, the dictionary distance for Turbine~5 is \blue{roughly} twice as large as \blue{the corresponding distance} of the healthy turbines prior to the replacement of the faulty bearing.
The two sparse coding algorithms \blue{we} investigated result in similar dictionary distance trends.
Note that in these numerical experiments, we use a sparse dataset with \blue{approximately} 2.56 seconds of recorded signal per day.
In an online monitoring implementation of this method, there would be significantly more data per time unit and a faster effective learning rate.
Thus, \blue{to achieve online processing of} the step length parameter $\eta$, \eqn{probreprSderSolvsimplr} should be lowered to avoid short-term overfitting of the propagated dictionary to different healthy operational states.

\subsection{ROC analysis}

We perform a basic receiver operating characteristic (ROC) analysis to study the usefulness of the dictionary distance measure as a condition indicator.
ROC curves are commonly used to asses the efficiency of condition indicators for diagnostics purposes~\cite{Fawcett2006}, and \blue{they} are used more generally as a method for classifier evaluation and selection.
A ROC curve illustrates the relationship between the true positive rate (TPR) and the false positive rate (FPR).
Each point on the curve corresponds to different parameters of the classifier model, for example, a threshold value of a condition indicator.
\red{This threshold value can be set up by an expert or an autonomous agent based on the performance of a population of turbines.
Thus,} the ROC curve describes the expected TPR and FPR for varying threshold values.

\begin{figure}[!h]
	\centering
	\begin{subfigure}{.4\textwidth}
		\centering
		\includegraphics[width=.9\linewidth]{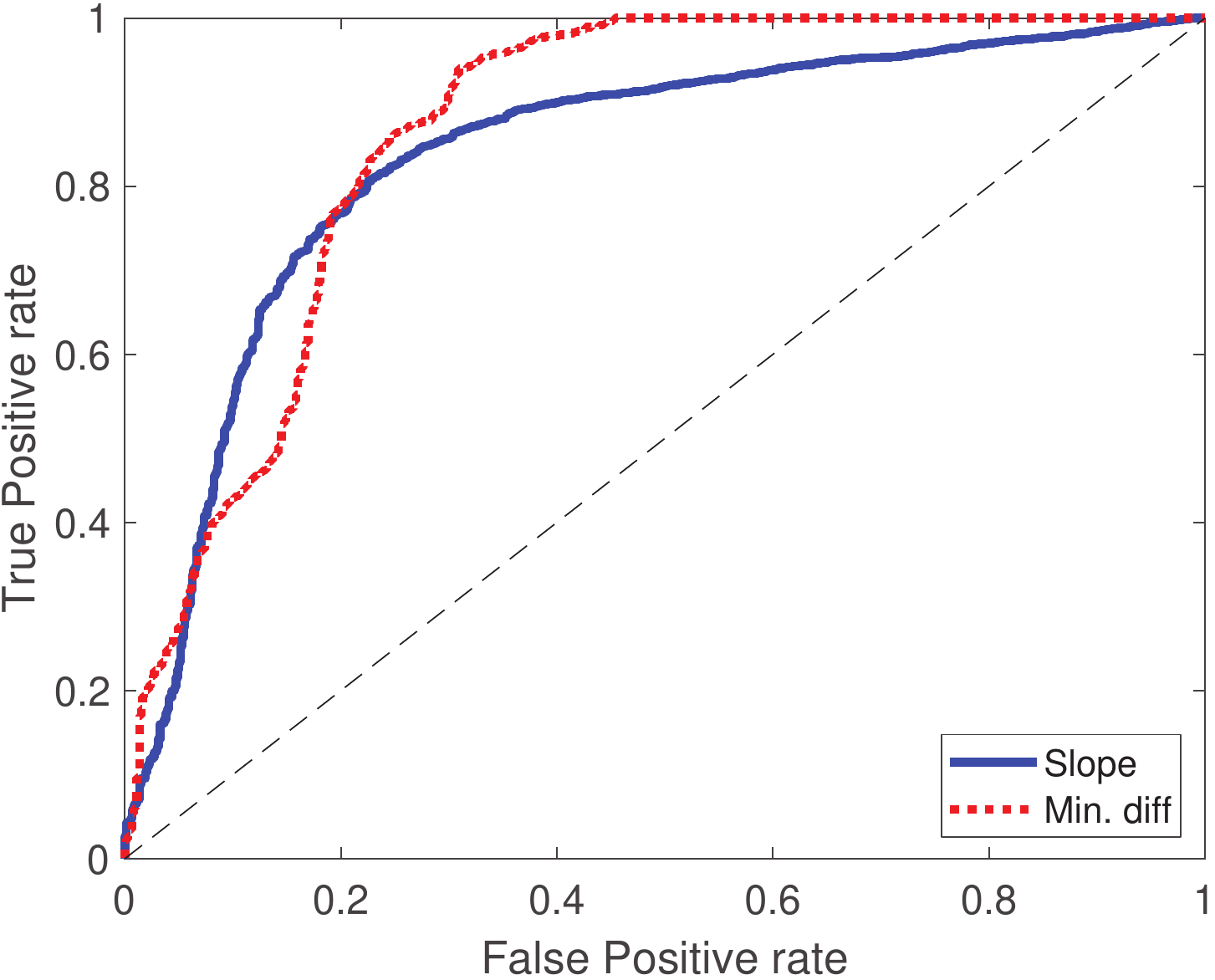}
		\caption{MP}
		\label{fig:ROCcurveMP}
	\end{subfigure}%
	\begin{subfigure}{.4\textwidth}
		\centering
		\includegraphics[width=.9\linewidth]{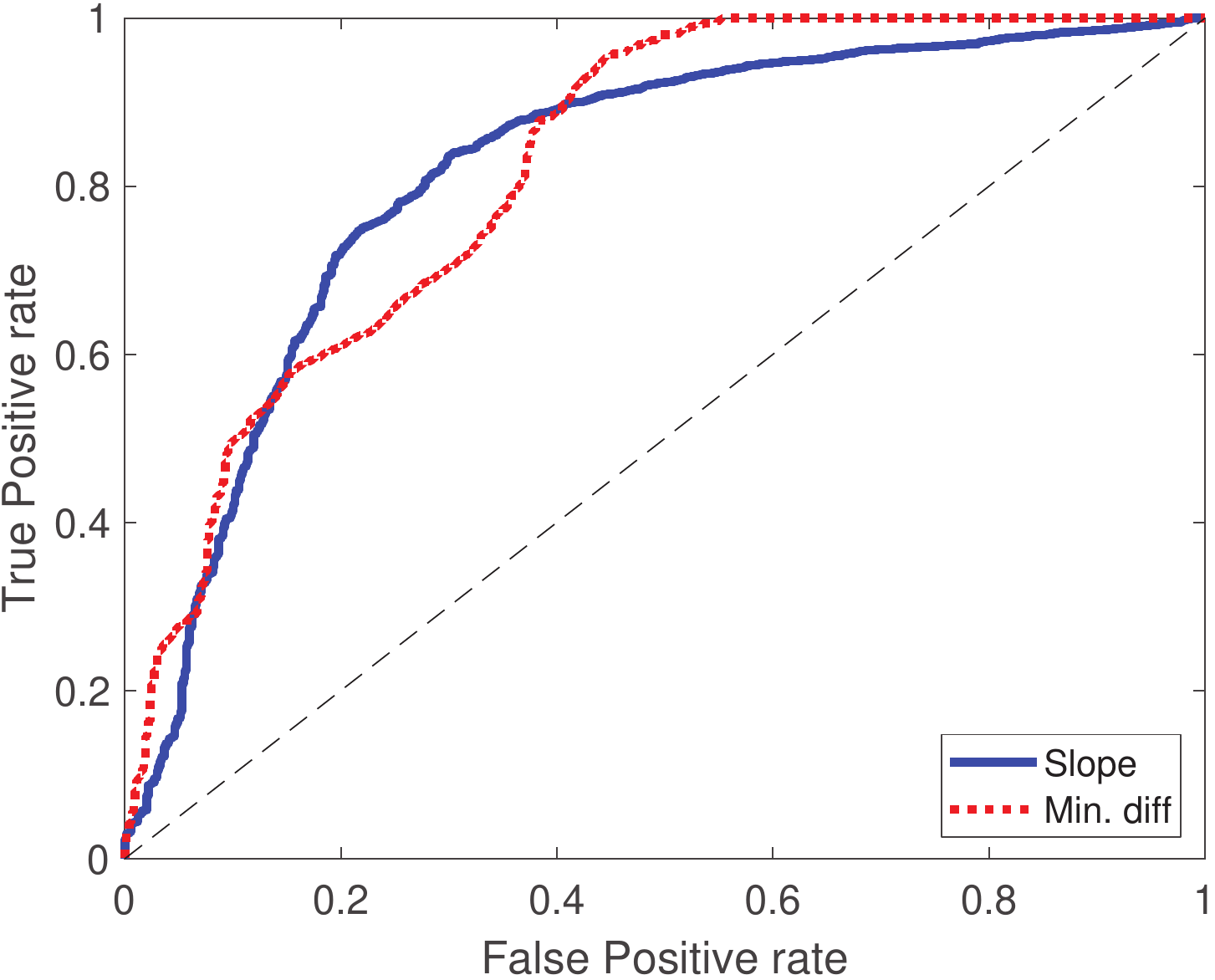}
		\caption{OMP}
		\label{fig:ROCcurveOMP}
	\end{subfigure}
	\caption{
		ROC curves based on the dictionary distances between \blue{the} baseline dictionaries and propagated dictionaries for the six turbines using the MP (a) and OMP (b) sparse coding algorithms.
		The curves result from a threshold on the rate of change of the distance over time (blue) and the minimum difference of \blue{the} dictionary distances in the population (red).
	}
	\label{fig:ROCcurve}
\end{figure}

\fig{ROCcurve} shows the ROC curves for two indicators based on the dictionary distance of the two sparse coding algorithms presented in \fig{Dict-distboth}.
One of the ROC curves is based on the slope of the dictionary distance versus time, and the other ROC curve is based on the minimum difference between the distance for one turbine compared to the distances for the other five turbines.
The slope-based indicator gives a balanced ratio of the number of true positives to false positives.
The minimum difference indicator is skewed, which means that the indicator makes positive classifications with weak confidence since
it classifies all positives correctly for false positive rates above \blue{approximately} one half.
The indicators based on the MP algorithm cover a larger area than the results obtained with the OMP algorithm.
This \blue{behavior} corresponds to higher classification accuracy at lower computational cost. 
Label B in \fig{Dict-distboth} marks the time of the bearing replacement in Turbine~5, while label A indicates the resolution of a suspected electrical issue \blue{that was introduced during the} installation of the sensor system in Turbine~2.
In the ROC analysis, we consider the data from Turbine~1, Turbine~3, Turbine~4, Turbine~6 and the data from Turbine~2 after time A as data that correspond to healthy states of operation.
Furthermore, the data from Turbine~5 after the gearbox replacement at C \blue{are} also considered as data that correspond to healthy states of operation.
The data from Turbine~5 before C and the data from Turbine~2 before A are considered as data that correspond to faulty states of operation.
The classification results used in the calculation of the TPR and FPR are defined by varying the threshold values of the slope-based and minimum difference indicators.

\section{Discussion}
\label{sec:discussion}

\subsection{Generalization of \blue{a} learned dictionary}

A working hypothesis that motivated this study is that condition monitoring signals from healthy turbines of the same type \blue{that} operate in similar environments \blue{with} nearby locations should be similar to some degree.
Thus, a baseline signal model learned from one healthy turbine or a set of turbines could be useful for monitoring other turbines.
We investigate this idea with vibration data recorded from accelerometers located at the same position in six wind turbines.
We find that the dictionaries learned for the different turbines have remarkable similarities, see \fig{TrainDict}.
Furthermore, a dictionary learned from one turbine is successfully applied to the other two turbines, see \fig{FidWD}.
This \blue{application cannot be made when one is} using an arbitrary dictionary learned from another bearing vibration dataset; see \fig{FidBDC}.
Thus, we conclude that a dictionary-based sparse signal model learned from a sensor in one turbine can \blue{be generalized} to the corresponding signal in another turbine, which \blue{permits} further studies in this direction.

For example, using data from a \blue{larger} population of turbines, it is possible to investigate whether there are some signal components or atoms that are common for all healthy turbines in a wind farm.
By learning dictionaries for a \blue{larger} population of wind turbines, one could also create a repository of dictionaries and dictionary elements, which would enable comparisons of dictionaries learned from similar turbines in different wind farms.
\blue{In this way,} the initial dictionary implemented in the condition monitoring systems for new wind turbines could be selected based on the typical features of healthy turbines.
Similarly, dictionaries and atoms learned from confirmed faulty turbines can be stored and potentially used for the diagnosis of similar faults in other turbines of the same type.

\subsection{Selection of \blue{a} baseline dictionary}

The baseline dictionary for each turbine considered here \blue{is} learned from vibration data recorded during a period of time when there is no known fault.
However, when taking a new turbine into service, it is not certain that the turbine \blue{will operate without problems or} that no faults were introduced during the installation of the wind turbine components and condition monitoring system.
Thus, if the baseline dictionary is trained \blue{by} starting from a randomized dictionary, it may not be possible to identify a fault that is already present in the turbine from \blue{its initial operation}.

The results presented in \fig{FidWD} show that the difference in fidelity when using a baseline dictionary learned from the turbine itself, or a baseline dictionary learned from \blue{a} similar turbine is small.
Thus, an alternative to using a randomized initial dictionary is to further investigate the possibility \blue{of using} a baseline dictionary learned from similar turbines that are known to operate in healthy conditions.
Using such a baseline dictionary, it could be possible to identify abnormal conditions \blue{that appear} when a turbine is taken into service.

\subsection{Selection of \blue{the} condition indicator}

In general, a condition indicator is a quantitative measure of the performance or operational condition of a machine, and a feature is a measurable characteristic used when modeling a signal or set of data.
Conventional condition indicators, such as the RMS and the energy within certain (kinematically determined) frequency bands, are successfully used as features both in conventional and machine learning approaches to condition monitoring.
However, this approach typically requires human expertise to select and customize indicators for each particular application and machine type, which is costly.
Furthermore, \blue{faults with unexpected characteristics can be difficult to detect with indicators engineered for particular purposes.
These facts} motivate investigations of unsupervised learning approaches \blue{as in} the present study, which can complement and potentially replace manually defined condition indicators in some applications.

For example, the absolute value or the
rate of change of the dictionary distance could be used as a condition indicator with a threshold level defining the allowed drift away from a baseline dictionary.
However, these methods require further testing with data from a larger population of wind turbines to determine the appropriate threshold value(s) and \blue{the} expected true and false positive rates.
Alternatively, dictionary distances can be used as scores in an unsupervised anomaly detection or ranking algorithm.
\blue{Thus,} unsupervised feature learning \blue{methods} such as dictionary learning could pave the way \blue{to} the development of unsupervised anomaly detection systems.

\subsection{Selection of sparse coding algorithm}

In this work the sparse representations are generated with Matching Pursuit (MP) and Orthogonal Matching Pursuit (OMP), which are both greedy sparse coding algorithms.
In both cases the same probabilistic gradient ascent algorithm is used for dictionary learning.
The OMP algorithm enforces \blue{the} orthogonality of the terms in \eqn{genmodel} by updating the weights of the previously selected terms for each consecutive term added to the sparse signal representation.
This \blue{procedure} results in higher computational cost and fidelity (given one particular dictionary) compared to the MP algorithm.
However, when \blue{they are} used in combination with dictionary learning, the fidelities achieved with the two methods can be comparable.

\fig{FidWD} shows that the difference in fidelity obtained with the two algorithms is negligible.
A decrease in the fidelity of the signal model for Turbine~5 when the HSS bearing is replaced at the beginning of the first year of operation is observed in both cases.
The update of the propagated dictionary depends on the selected sparse coding algorithm, which consequently affects the calculated dictionary distance.
\fig{Dict-distboth} shows that the dictionary distance distribution is wider under normal operating conditions in the case of the OMP algorithm compared to the MP algorithm.
Considering the higher computational cost of the OMP algorithm, these results indicate that there is no evident benefit \blue{to} using OMP for the calculation of \blue{dictionary-learning-based} condition indicators and that MP-based indicators (for unknown reasons) may be beneficial for anomaly detection.

The computational cost of the sparse coding algorithm is significant, even if the implementations of the MP and OMP algorithms used here have been optimized for efficiency.
Thus, an interesting direction for future work is to investigate alternative methods for unsupervised feature learning, such as learning of co-sparse analysis operators \cite{Seibert2016}, where the inverse problem addressed here is avoided.

\section{Conclusion}
\label{sec:conclusion}

This work focuses on the monitoring of rolling element bearings in wind turbines using an unsupervised dictionary learning approach and
real--world wind turbine vibration data that \blue{have been} made publicly available.
The results presented above \blue{demonstrate that an indicator based on the dictionary distance is useful for the} condition monitoring of wind turbines \blue{and serves} as a complement to \blue{currently used methods.}
In the case of Turbine~5 considered above, it is not known when the issue(s) leading to the bearing and gearbox replacements first appeared.
There was a sudden increase \blue{in} the enveloped signal from the HSS axial sensor
after \blue{approximately} one year of operation, which motivated the bearing replacement.
\blue{This replacement was} followed by a gearbox replacement \blue{approximately} one year later.
\fig{Dict-distboth} suggests that the abnormal behavior of Turbine~5 could have been detected several months earlier using dictionary learning, which is an improvement in terms of maintenance planning and reducing the risk of costly failures.
However, further tests are required to understand the strengths and weaknesses of a dictionary learning approach in a realistic large-scale monitoring application.
For example, it is not understood whether the long-term drifts away from the baseline dictionaries observed in \fig{Dict-distboth} are related to mechanical wear of the turbines, or the greedy approximation of the NP hard dictionary learning problem.
Further testing requires acquisition and processing of condition monitoring data from a larger population of turbines, including documented faults and maintenance activities, which is the next step \blue{but is} beyond the scope of the project reported here.

\section*{Acknowledgments}

We thank Per-Erik Larsson, Stephan Schnabel, Allan Thomson and Joe Erskine for discussions that helped us improve the manuscript, and \blue{we thank} Karl Skretting for contributing the idea to use a unique seed dictionary.
This work is partially supported by SKF through their University Technology Center at LTU.
\blue{The contribution of Fredrik Sandin} is funded by the Kempe Foundations under contract Gunnar {\"O}quist Fellow.
The research leading to these results has received funding from the People Programme (Marie Curie Actions) of the European Union's Seventh Framework Programme FP7/2007-2013/ under REA grant agreement number 612603 and the Swedish Foundation for International Cooperation in Research and Higher Education (STINT), grant number IG2011-2025.

\appendix

\section{Data}
\label{app:data}

The raw time-domain vibration signals and speed data from the six wind turbines that are analyzed for the first time in this article are publicly available at the permanent link \url{http://urn.kb.se/resolve?urn=urn:nbn:se:ltu:diva-70730}.
The sampling rate is 12.8 \red{kHz,} and the signal segments are 1.28 seconds long (16384 samples per segment). 
Signal segments are recorded with an interval of approximately 12 hours over a period of 46 consecutive months for each turbine.
Over this time period, bearing and gearbox faults \blue{appeared} in one of the six turbines, as described in Sections 3 and 4.







\section*{References}
\bibliographystyle{elsarticle-num}




\end{document}